\documentclass[10pt,journal,compsoc]{IEEEtran}

\ifCLASSOPTIONcompsoc
  \usepackage[nocompress]{cite}
\else
  \usepackage{cite}
\fi

\ifCLASSINFOpdf
\else
\fi

\usepackage{amsmath}
\usepackage{booktabs} 
\usepackage{graphicx}
\usepackage{multirow}
\usepackage{multicol}
\usepackage{bm}
\usepackage{subcaption}
\usepackage{amssymb}
\usepackage{rotating}
\usepackage{tikz}
\usepackage{placeins}

\begin{document}

\title{As-Rigid-As-Possible Deformation of Gaussian Radiance Fields}

\author{Xinhao~Tong,
        Tianjia~Shao,
        Yanlin~Weng,
        Yin~Yang,        and~Kun~Zhou,~\IEEEmembership{Fellow,~IEEE}
\IEEEcompsocitemizethanks{\IEEEcompsocthanksitem Xinhao Tong, Tianjia Shao, Yanlin Weng, Kun Zhou are with the State Key Lab of CAD\&CG, Zhejiang University, Hangzhou 310058, China. \protect\\ E-mail: \{xinhaot, tjshao\}@zju.edu.cn, weng@cad.zju.edu.cn, kunzhou@acm.org. Yanlin Weng is the corresponding author.
\IEEEcompsocthanksitem Yin Yang is with Kahlert School of Computing, University of Utah, USA. E-mail: yangzzzy@gmail.com.
\IEEEcompsocthanksitem The project is supported by NSF China (No. 62172357 \& 62421003 \& 62322209), NSF 2301040, the XPLORER PRIZE, and the gift from Adobe Research.
}

\thanks{Manuscript received April 19, 2005; revised August 26, 2015.}}

\markboth{Journal of \LaTeX\ Class Files,~Vol.~14, No.~8, August~2015}%
{Shell \MakeLowercase{\textit{et al.}}: Bare Demo of IEEEtran.cls for Computer Society Journals}

\IEEEtitleabstractindextext{%
\begin{abstract}
3D Gaussian Splatting (3DGS) models radiance fields as sparsely distributed 3D Gaussians, providing a compelling solution to novel view synthesis at high resolutions and real-time frame rates. However, deforming objects represented by 3D Gaussians remains a challenging task. Existing methods deform a 3DGS object by editing Gaussians geometrically. These approaches ignore the fact that it is the radiance field that rasterizes and renders the final image. The inconsistency between the deformed 3D Gaussians and the desired radiance field inevitably leads to artifacts in the final results. In this paper, we propose an interactive method for as-rigid-as-possible (ARAP) deformation of the Gaussian radiance fields. Specifically, after performing geometric edits on the Gaussians, we further optimize Gaussians to ensure its rasterization yields a similar result as the deformed radiance field. To facilitate this objective, we design radial features to mathematically describe the radial difference before and after the deformation, which are densely sampled across the radiance field. Additionally, we propose an adaptive anisotropic spatial low-pass filter to prevent aliasing issues during sampling and to preserve the field with the varying non-uniform sampling intervals. Users can interactively employ this tool to achieve large-scale ARAP deformations of the radiance field. Since our method maintains the consistency of the Gaussian radiance field before and after deformation, it avoids artifacts that are common in existing 3DGS deformation frameworks. Meanwhile, our method keeps the high quality and efficiency of 3DGS in rendering.

\end{abstract}

}

\maketitle
\IEEEdisplaynontitleabstractindextext
\IEEEpeerreviewmaketitle

\IEEEraisesectionheading{\section{Introduction}\label{sec:introduction}}

\IEEEPARstart{D}{eforming} 3D objects is a longstanding topic in computer graphics, underpinning tasks such as 3D modeling and animation. It has been extensively studied for models with an explicit discretization e.g., meshes, point clouds, or voxel grids~\cite{Yu04,Zhou05,Huang06,SorkineHornung06}.
Recently, inspired by the success of implicit representations for novel view synthesis like NeRF~\cite{MildenhallSTBRN20,BarronMTHMS21,BarronMVSH22} and NGP~\cite{MullerESK22}, many efforts have also explored deformable objects within these frameworks. 3D Gaussian Splatting (3DGS) combines an explicit Gaussian-based representation with differentiable rasterization~\cite{KerblKLD23}. It offers more efficient training and rendering while capturing high-frequency details than fully implicit models.

The explicit nature of 3DGS suggests the possibility of deforming 3D objects via direct geometric edits of Gaussian kernels. For instance, SuGaR~\cite{Gudon23} binds Gaussians to an extracted surface mesh. By deforming the mesh, the associated Gaussians are altered, enabling scene deformation. SC-GS~\cite{Huang23} learns a set of sparse control points from videos of dynamic objects, and manipulates those points to modify the position and rotation of Gaussians. GaMeS~\cite{abs-2402-01459} binds each flat Gaussian to a triangular patch, enabling the editing of each Gaussian's position, rotation, and scale by manipulating the vertex positions of the triangles. GaussianMesh~\cite{Gao24} initializes Gaussians on a mesh extracted by NeuS2~\cite{WangHHDTL23}. It incorporates face-splitting operations and Gaussian size constraints during reconstruction to mitigate artifacts. GaussianFrosting~\cite{GSFrosting} constrains Gaussians within a frosting layer to preserve surface details, such as hair-like features, and edits them by altering vertex positions within partitioned cells. On the other hand, rasterizing those deformed Gaussians often yields artifacts, especially when the model undergoes nonlinear and localized deformation. This issue becomes more severe at the boundary of the model. 
The underlying reason is \emph{the mismatch between deformed Gaussians and the deformed radiance field}. Concretely, a deformed 3DGS encodes a deformation field with piece-wise deforming Gaussian kernels. On the other hand, the actual radiance field is expected to have a spatially varying deformation over the scene --- different positions possess different local deformations in the deformed radiance field, which does not perfectly align with deformed Gaussians. This discrepancy becomes more visible under large deformations, and the resultant image-space rendering turns out less satisfactory. In other words, a high-quality 3DGS-based deformable model should concern the deformation of the radiance field induced by deformed Gaussian kernels, which is a critical overlook from existing methods.

\begin{figure}[t]
    \centering
    \begin{subfigure}[b]{0.48\textwidth}
        \centering
        \includegraphics[width=\textwidth]{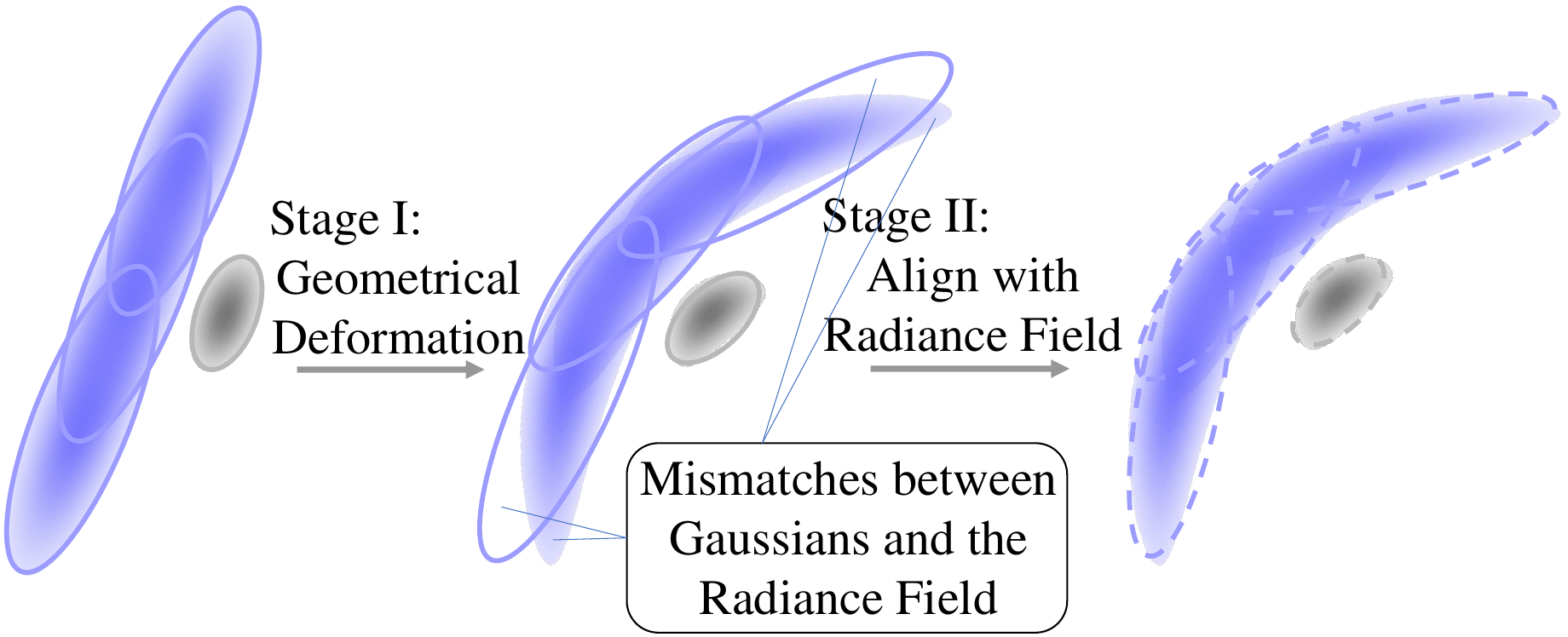}  
    \end{subfigure}
    \caption{\textbf{A conceptual diagram.} In the conceptual diagram, we illustrate our two-stage approach. In the leftmost figure, the initial state of the Gaussians and the radiance field is shown. The solid contours represent the ellipsoids of the Gaussians, while the gradient colors indicate the radiance field. The middle figure depicts the result after geometric deformation, where mismatches between the Gaussians and the radiance field can be observed, especially for elongated Gaussians undergoing large deformations. The rightmost figure shows the result of optimizing the Gaussians to align with the radiance field, significantly reducing the aforementioned mismatches. The dashed ellipses represent the optimized Gaussians.}
    \label{fig:conceptual}
\end{figure}

We propose a method enabling interactive As-Rigid-As-Possible (ARAP) deformations over a radiance field represented by Gaussian Splatting. Our approach accommodates both 3DGS and flat Gaussian (in GaMeS~\cite{abs-2402-01459}), and it supports large-scale ARAP deformations unnecessitating an explicit surface reconstruction. In addition to geometric deformation directly applied at Gaussians, we take radiance field consistency into account. Specifically, we use the embedded deformation~\cite{SumnerSP07} to model the geometric transformation of Gaussians driven by an underlying deformation graph. After that, we extract \emph{radial features} across the radiance field associated with the deformed Gaussians and minimize the feature errors between the radiance field and Gaussian rasterization. To address the potential aliasing issue from spatial sampling variations, we propose an adaptive anisotropic low-pass filter that compensates for changes in sampling densities. This approach preserves the radiance field's consistency during deformation and effectively removes artifacts while maintaining the efficiency and quality of 3DGS rendering. Fig. \ref{fig:conceptual} illustrates the core idea of our approach.
Experiments on synthetic and real-world data demonstrate that our method achieves higher rendering quality for complex, nonlinear, and large deformations. Quantitative experiments further validate our superior benchmark. Please refer to the supplementary video for animated results of interactive deformation processes and novel-view synthesis. The main contributions of this paper are as follows:


\begin{itemize}
    \item To the best of our knowledge, we are the first to optimize deformed Gaussians based on the radiance field they represent, which closes the gap between per-Gaussian deformation and deformation over the radiance field. 
    \item We propose an interactive deformation method for the Gaussian radiance fields. This method does not require surface extraction and supports large-scale ARAP deformations.
    \item We propose an adaptive anisotropic spatial low-pass filter that can prevent aliasing effects under changing non-uniform sampling intervals during deformation.
\end{itemize}



\section{Related Work}

\subsection{Radiance Fields}
In recent years, radiance field based approaches have achieved promising results in 3D reconstruction and novel view synthesis. Among these, the neural radiance field (NeRF)~\cite{MildenhallSTBRN20} stands as a foundational work. NeRF takes multi-view images as input and trains an MLP network to represent color and opacity information in 3D space, achieving good results in novel view synthesis through volume rendering techniques. Subsequent works on implicit radiance fields have enhanced NeRF in various aspects, including rendering speed, training efficiency, reconstruction quality, and the ability to represent dynamic scenes. Mip-NeRF~\cite{BarronMTHMS21} enhances the rendering quality of NeRF by employing multi-scale representations to achieve anti-aliasing. PlenOctrees~\cite{YuLT0NK21} and Instant-NGP~\cite{MullerESK22} integrate implicit radiance fields with octrees and hash grids, respectively, significantly improving rendering speed while also enhancing reconstruction quality. PointNeRF~\cite{XuXPBSSN22} combines point clouds with neural features, enabling the reconstruction of 3D scenes through training while maintaining high rendering efficiency and detailed expression capabilities. D-NeRF~\cite{PumarolaCPM21} incorporates time as an additional variable encoded into the network, enabling the reconstruction of dynamic 3D scenes.

3D Gaussian Splatting (3DGS)~\cite{KerblKLD23} combines explicit Gaussian representations with differentiable rasterization to achieve high quality and efficiency in novel view synthesis. It utilizes 3D Gaussian kernels with opacity and represents the anisotropic color of each Gaussian using spherical harmonics (SH). The method optimizes the Gaussians through a differentiable rasterization rendering technique, resulting in high training and rendering speeds. In recent months, there are some advanced works of 3DGS. Mip-Splatting~\cite{Mipsplatting} introduces a 3D low-pass filter to achieve anti-aliasing, thereby enhancing the rendering quality of Gaussians at varying distances. The algorithms of 4DGS~\cite{4DGS}, Dynamic Gaussian~\cite{LuitenKLR24}, and Deformable Gaussian~\cite{DeformableGS} incorporate the time variable, endowing Gaussians with the capability to represent dynamic scenes. 

\begin{figure*}[h!]
    \centering

    \begin{subfigure}[b]{0.99\textwidth}
        \centering
        \includegraphics[width=\textwidth]{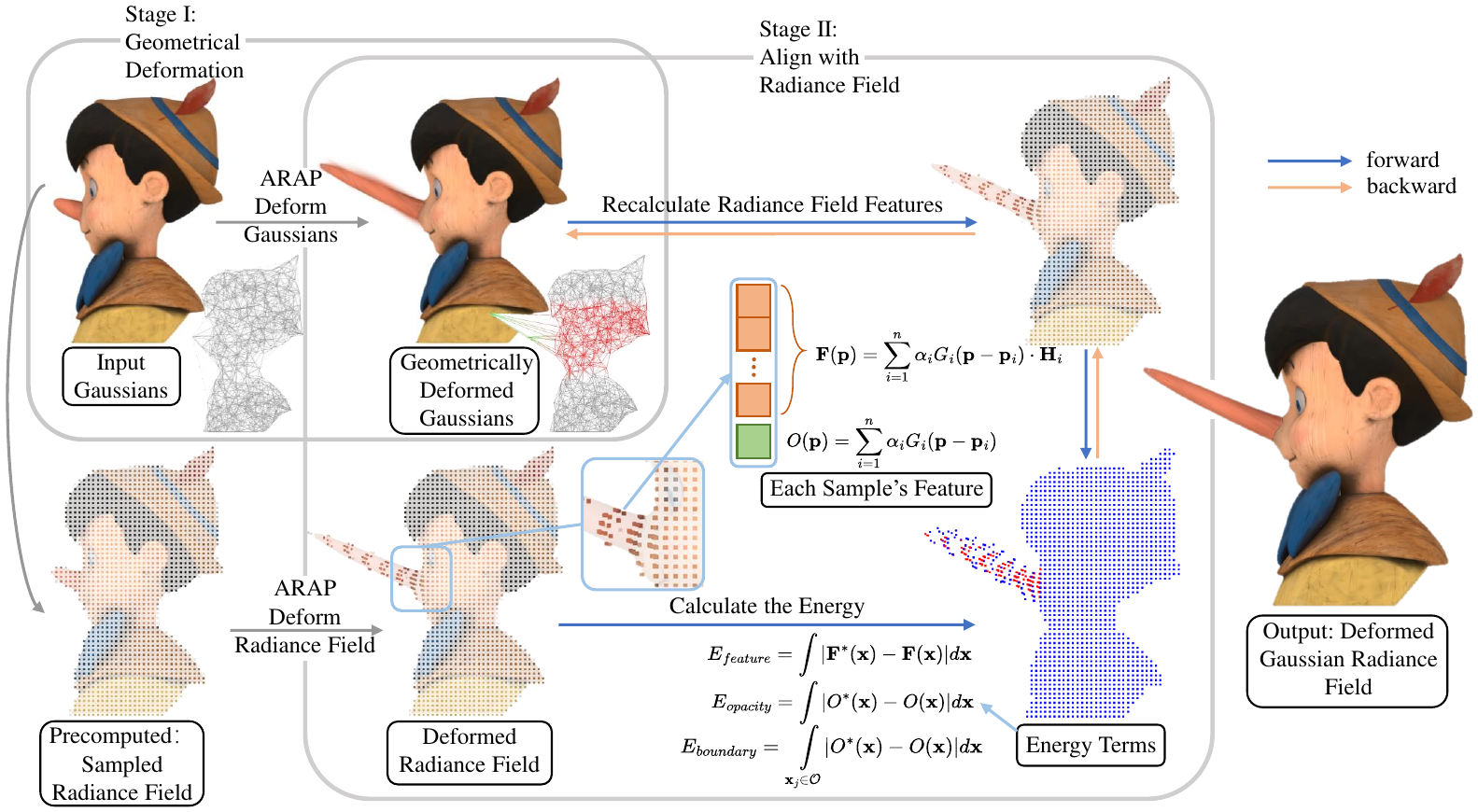}  
    \end{subfigure}

    \caption{\textbf{A general overview of the pipeline.} Our pipeline for ARAP deformation of Gaussian radiance fields is illustrated in the figure. Given a Gaussian object as input, we first perform a geometric deformation on the Gaussians using the embedded deformation. In the second stage, we treat the object represented by the Gaussians as a radiance field. Based on the discrepancy between the deformed radiance field and the radiance field expressed by the geometrically deformed Gaussians, we design several energy terms. We optimize the Gaussians to minimize these energy terms, ensuring that the optimized Gaussians conform to the deformation of the radiance field. Visually, this approach eliminates artifacts that arise from merely performing geometric deformation on the Gaussians.}
    \label{fig:pipeline}
\end{figure*}

\subsection{3D Shape Deformation}
There have been numerous works in 3D mesh deformation such as mesh deformation based on Possion~\cite{Yu04} and Laplacian Coordinates~\cite{Zhou05, Huang06, SorkineHornung06}. Some works~\cite{SorkineA07} aim to perform as-rigid-as-possible (ARAP) deformation on shapes to preserve local geometric details. Embedded Deformation~\cite{SumnerSP07} embeds ARAP Deformation into space, enabling it to support editing of various geometric representations. 

In recent years, with the rise of implicit 3D representations exemplified by NeRF~\cite{MildenhallSTBRN20}, there has been a surge in works focusing on editing or deforming implicit 3D representations. CageNeRF~\cite{PengYLCGPZY22}, Deforming-NeRF~\cite{XuH22} and NeRFshop~\cite{JambonKKDLD23} utilize cage-based deforming and ray-bending techniques to edit or deform scenes represented by NeRF or Instant-NGP~\cite{MullerESK22}. NeRF-Editing~\cite{YuanSLMJ022} and Interactive-NeRF~\cite{YuanSLMJKG23} initially utilize NeuS~\cite{WangLLTKW21} to extract the implicit surface of the object and reconstruct the mesh and bounding cage. ARAP deformation constraints are then applied to the mesh, enabling ARAP deformation of objects represented by NeRF. NeuMesh~\cite{YangBZBZCZ22} and DE-NeRF~\cite{0009SL023} encode neural implicit fields onto a mesh, enabling deformation of the implicit field by deforming the mesh with neural features. Due to the inability of NeRF-based representations to simultaneously achieve high rendering quality and fast rendering speed, the aforementioned work exhibits suboptimal rendering performance when supporting interactive deformations.

With the introduction of 3D Gaussian Splatting~\cite{KerblKLD23}, recently, there has also been some works focusing on editing 3DGS~\cite{WuYZYCYG24}. SC-GS~\cite{Huang23} learns sparse control points from dynamic scenes and utilizes them to drive 3DGS. SuGaR~\cite{Gudon23} extracts explicit meshes from 3DGS and performs joint optimization on Gaussians and meshes. This allows it to edit 3DGS through edits on the mesh. GaussianMesh~\cite{Gao24}, utilizes NeuS2~\cite{WangHHDTL23} to extract meshes from objects and optimizes Gaussians bound to mesh faces, providing a method for performing ARAP deformation on the mesh to edit Gaussians. GaussianFrosting~\cite{GSFrosting}, after extracting the mesh, establishes a frosting layer around the mesh. This layer is an adaptive layer with variable thickness designed to capture hair-like details near the surface. During Gaussian optimization, the range of Gaussian positions is confined to the frosting layer. In deformation, prismatic cells are created within the frosting layer to enclose the Gaussians, and the positions and shapes of the Gaussians are edited based on the changes in the vertex positions of the prismatic cells. Mani-GS~\cite{Mani-GS} binds Gaussians to a given mesh for reconstruction, placing Gaussians in the local coordinate system of faces of mesh and driving the Gaussians through mesh editing. GSDeformer~\cite{GSDeformer} constructs a cage around the Gaussians and directly edits the Gaussians inside the cage by editing the vertices of the cage. GaMeS~\cite{abs-2402-01459} binds flat Gaussians to a triangle mesh or triangle soup, allowing modification of the Gaussians by editing the vertices' positions of the triangles. There are also some researches on physics-informed Gaussian Splatting~\cite{Xie23, abs-2401-15318, JiangYXLFWLLG0J24}. For example, PhysGaussian~\cite{Xie23} utilizes discrete particle clouds derived from 3DGS, and use them to perform physics-based deformation. 

All the above methods did not consider the mismatch between deformed Gaussians and the deformed radiance field, which will produce non-negligible artifacts under large deformations. 


\section{Methodology}\label{sec:method}
We aim to provide a high-quality ARAP deformation pipeline for a 3DGS radiance field and avoid artifacts that may arise from directly deforming Gaussians. As outlined in Fig.~\ref{fig:pipeline}, we employ a two-stage approach to achieve this goal. In the first stage, we utilize embedded deformation to apply geometric modifications to 3D Gaussians (which is similar to many existing methods). In the second stage, we seek a better-optimized configuration of each Gaussian to minimize the discrepancy between the deformed 3DGS and the deformed radiance field. While we choose ARAP deformation in this framework due to its popularity in graphics, our method is compatible with other types of deformations, either geometric-based or physics-based. 


\subsection{Stage I: Geometrical Deformation of Gaussians}\label{sec:stage1}
Embedded deformation~\cite{SumnerSP07} offers a versatile mechanism to apply ARAP deformation over a range of geometry representations such as meshes, point clouds, and particles. It is based on the concept of deformation graph. Specifically, it uses a set of affine transformations to encode a spatially varying ARAP deformation field. Each affine transformation is associated with a control node as a vertex in the deformation graph, which includes a $3 \times 3$ matrix $\bm{A}$ and a $3 \times 1$ translation vector $\bm{t}$. Let the position of the control point before and after deformation be $\bm{g}$ and $\bm{\tilde{g}}$, and $\bm{q}_l$ be the desired deformed position of the $l$-th control point (which is specified by the user). We seek for the optimal configuration of all the transformations to minimize three energies $E_{rot}$, $E_{reg}$, and $E_{con}$ defined as:
\begin{align}
         E_{rot}(\bm{A}) &= (\bm{c}_1 \cdot \bm{c}_2)^2 + (\bm{c}_1 \cdot \bm{c}_3)^2 + (\bm{c}_2 \cdot \bm{c}_3)^2 + \nonumber \\ 
         & (\bm{c}_1 \cdot \bm{c}_1 - 1)^2 + (\bm{c}_2 \cdot \bm{c}_2 - 1)^2 + (\bm{c}_3 \cdot \bm{c}_3 - 1)^2,
\end{align}
\begin{equation}
    E_{reg}(\bm{A})  = \sum_{k} \left\|\bm{A}(\bm{g}_k-\bm{g})+\bm{g}+\bm{t}-(\bm{g}_k+\bm{t}_k)\right\|_2^2,
\end{equation}
and
\begin{equation}
    E_{con} = \sum_{l} \|\tilde{\bm{g}}_l-\bm{q}_l\|_2^2.
\end{equation}
Here, $\bm{c}_1$, $\bm{c}_2$, $\bm{c}_3 \in\mathbb{R}^3$ are three column vectors of $\bm{A}$, and $E_{rot}$ measures how far is $\bm{A}$ from an ideal rigid rotation. The summation index $k$ in $E_{reg}$ iterates all the adjacent control nodes on the deformation graph. Clearly, $E_{reg}$ penalizes sharp and abrupt translational deformation of the control point. The last energy $E_{con}$ corresponds to how well the user's constraints are satisfied. During the deformation process, the total energy $E_{geo}$ is minimized to maintain the deformation as rigid as possible, which is defined as: 
\begin{equation}
    E_{geo} = \sum_{i}(w_{rot}E_{rot}(\bm{A}_i)+w_{reg}E_{reg}(\bm{A}_i)) + w_{con}E_{con}.
\end{equation}
Here $w_{rot} = 1$, $w_{reg} = 10$, and $w_{con} = 100$, which are the same as~\cite{SumnerSP07}. 

The resulting affine transformations are then used to displace of embedded geometries by from $\bm{v}_i$ to $\tilde{\bm{v}}_i$:
\begin{equation}
    \bm{\tilde{v}}_i = \sum_j w_j(\bm{v}_i)\big(\bm{A}_j(\bm{v}_i - \bm{g}_j) + \bm{g}_j + \bm{t}_j\big),
\end{equation}
where $w_j(\bm{v}_i) \geq 0$ is the weight or influence inherited from the $j$-th control node to $\bm{v}_i$.


We choose control points out of Gaussian centers following furthest point sampling (FPS). These control points serve as vertices in the deform graph, whose edge connections are constructed using the KNN algorithm. All the Gaussians can now be displaced and deformed (in an ARAP way) following the resultant deformation graph. To capture per-Gaussian deformation, we select six endpoints for each Gaussian along the three axes away from its center of twice the standard deviation. Those endpoints move along the Embedded Deformation, and the deformed Gaussian is reconstructed based on the deformed positions of endpoints. Let $\bm{Q} = [\bm{q}_1, \bm{q}_2, \bm{q}_3, \bm{q}_4, \bm{q}_5, \bm{q}_6] \in \mathbb{R}^{3\times 6}$ and $\bm{P} = [\bm{p}_1, \bm{p}_2, \bm{p}_3, \bm{p}_4, \bm{p}_5, \bm{p}_6] \in \mathbb{R}^{3\times 6}$ encode the original and deformed position of endpoints. The best-fitting local deformation of the Gaussian is estimated as:
\begin{equation}\label{eq:local_transformation}
    \bm{M} = \left(\bm{P}\bm{Q}^\top\right) \left(\bm{Q}\bm{Q}^\top\right)^{-1}.
\end{equation}
We use polar decomposition to obtain the deformed Gaussian's rotation $\bm{r}$ and scaling $\bm{s}$. The spherical harmonics of each Gaussian are also subjected to corresponding rotations. 


\begin{figure}[t]
    \centering
    \begin{subfigure}[b]{0.48\textwidth}
        \centering
        \includegraphics[width=\textwidth]{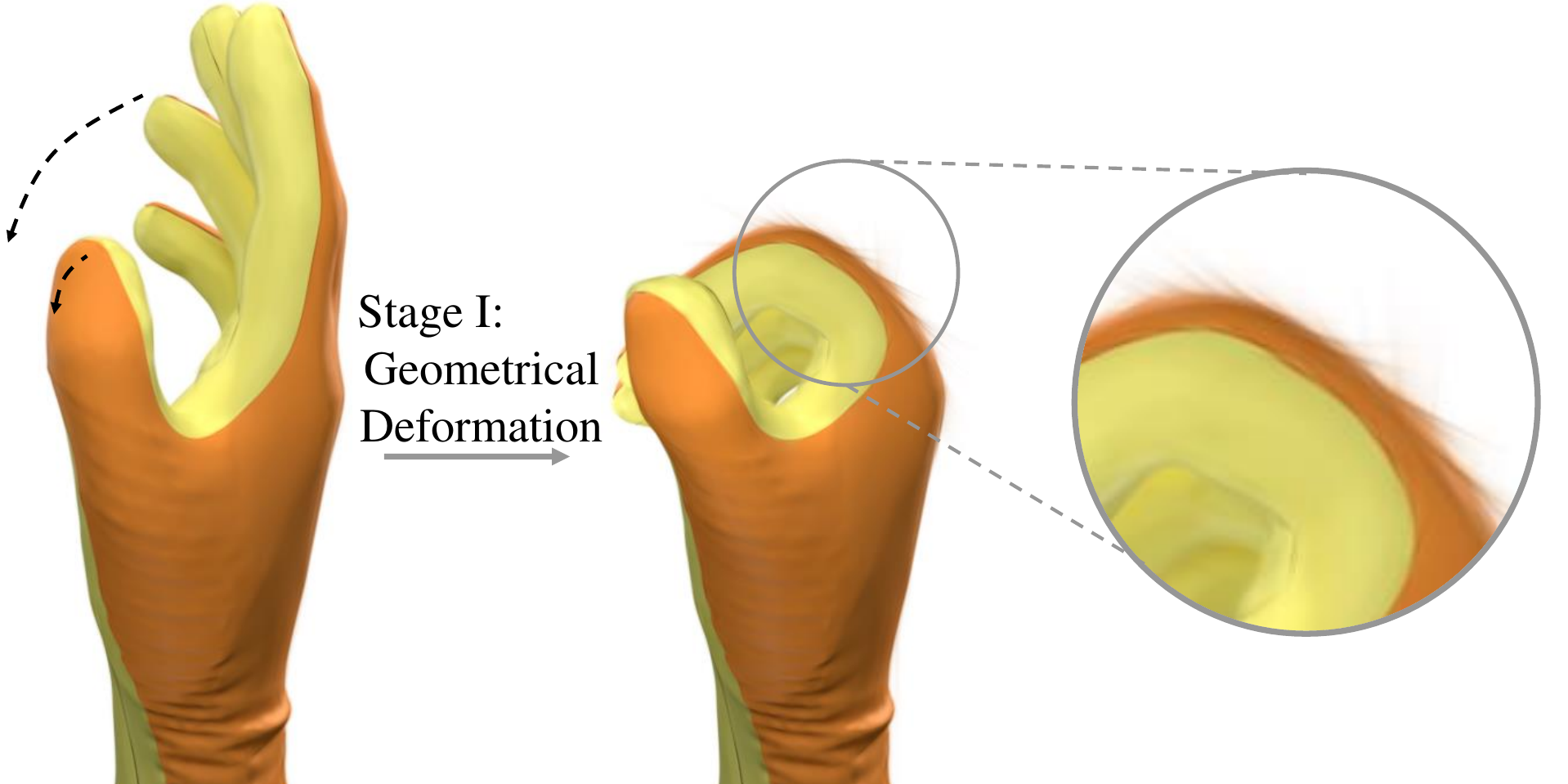}  
    \end{subfigure}
    \caption{\textbf{Artifacts in the results of Stage I.} After the geometrical deformation in Stage I, visual artifacts often appear, as shown in the figure.}
    \label{fig:artifact}
\end{figure}

\subsection{Stage II: Align 3DGS with Radiance Field}\label{sec:stage2}
Deformed Gaussians, when rasterized from novel views, often generate visual artifacts under large-scale deformations, as shown in Fig. \ref{fig:artifact}. Suppressing sharp stretching may mitigate this issue~\cite{Xie23} at the cost of the significantly increased number of Gaussians. The actual reason behind this issue is that rasterizing deformed 3DGS does not always align with the rendering obtained from the deformed radiance field. Our second stage aims to measure this discrepancy and finetune the configuration of each Gaussian to make sure it faithfully captures the field-wise deformation.   

\begin{figure*}[ht!]
    \centering
    \begin{subfigure}[b]{0.16\textwidth}
        \centering
        Input \& Deform
    \end{subfigure}
    \hfill
    \begin{subfigure}[b]{0.16\textwidth}
        \centering
        Baseline
    \end{subfigure}
    \hfill
    \begin{subfigure}[b]{0.16\textwidth}
        \centering
        Ours
    \end{subfigure}
    \hfill
    \begin{subfigure}[b]{0.16\textwidth}
        \centering
        Input \& Deform
    \end{subfigure}
    \hfill
    \begin{subfigure}[b]{0.16\textwidth}
        \centering
        Baseline
    \end{subfigure}
    \hfill
    \begin{subfigure}[b]{0.16\textwidth}
        \centering
        Ours
    \end{subfigure}

    \begin{subfigure}[b]{0.49\textwidth}
        \centering
        \includegraphics[width=\textwidth]{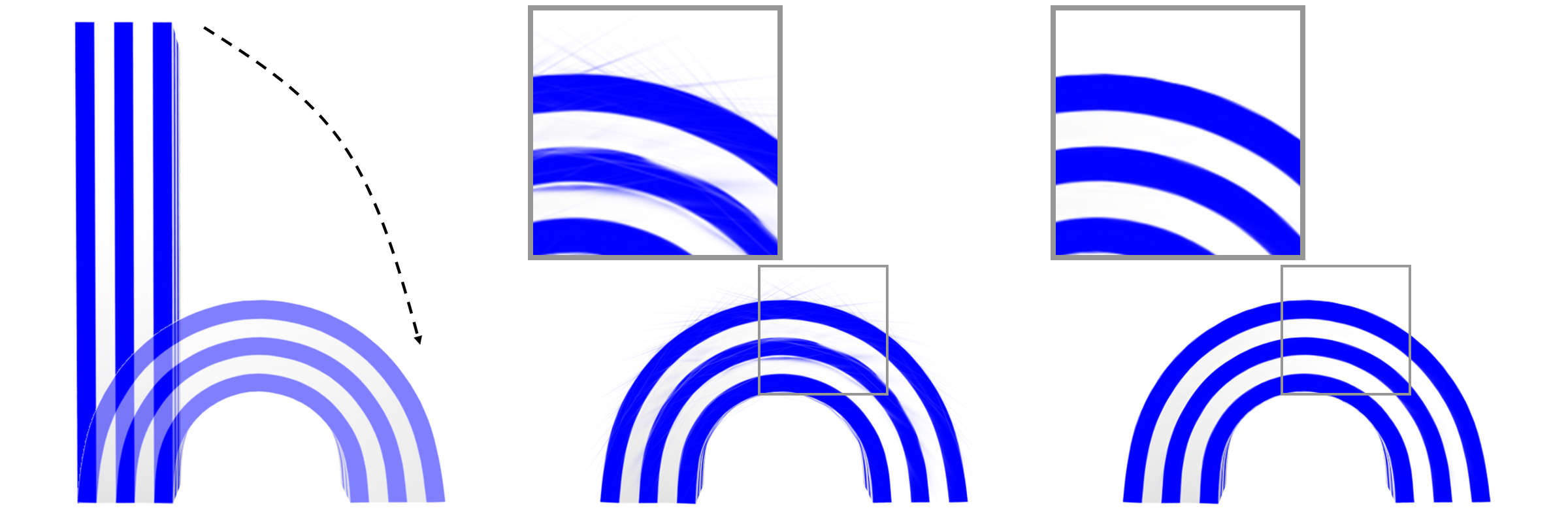}  
    \end{subfigure}
    \hfill
    \begin{subfigure}[b]{0.49\textwidth}
        \centering
        \includegraphics[width=\textwidth]{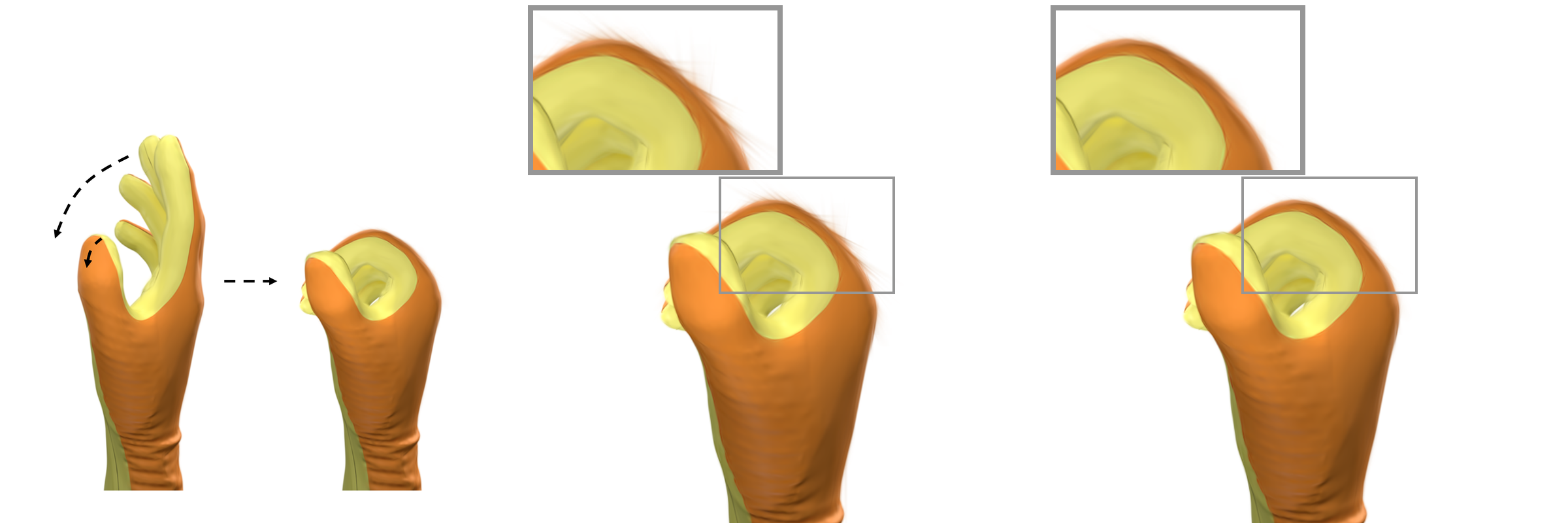}  
    \end{subfigure}

    \begin{subfigure}[b]{0.49\textwidth}
        \centering
        (a) Stripes
    \end{subfigure}
    \hfill
    \begin{subfigure}[b]{0.49\textwidth}
        \centering
        (b) Glove
    \end{subfigure}

    \begin{subfigure}[b]{0.49\textwidth}
        \centering
        \includegraphics[width=\textwidth]{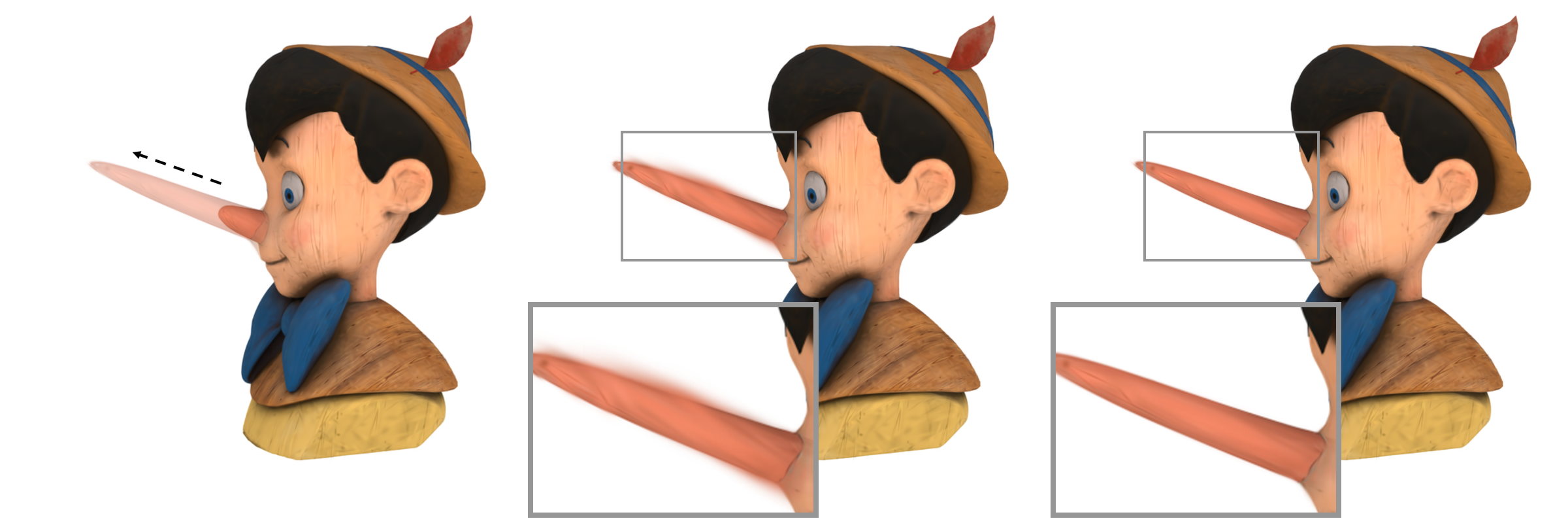}  
    \end{subfigure}
    \hfill
    \begin{subfigure}[b]{0.49\textwidth}
        \centering
        \includegraphics[width=\textwidth]{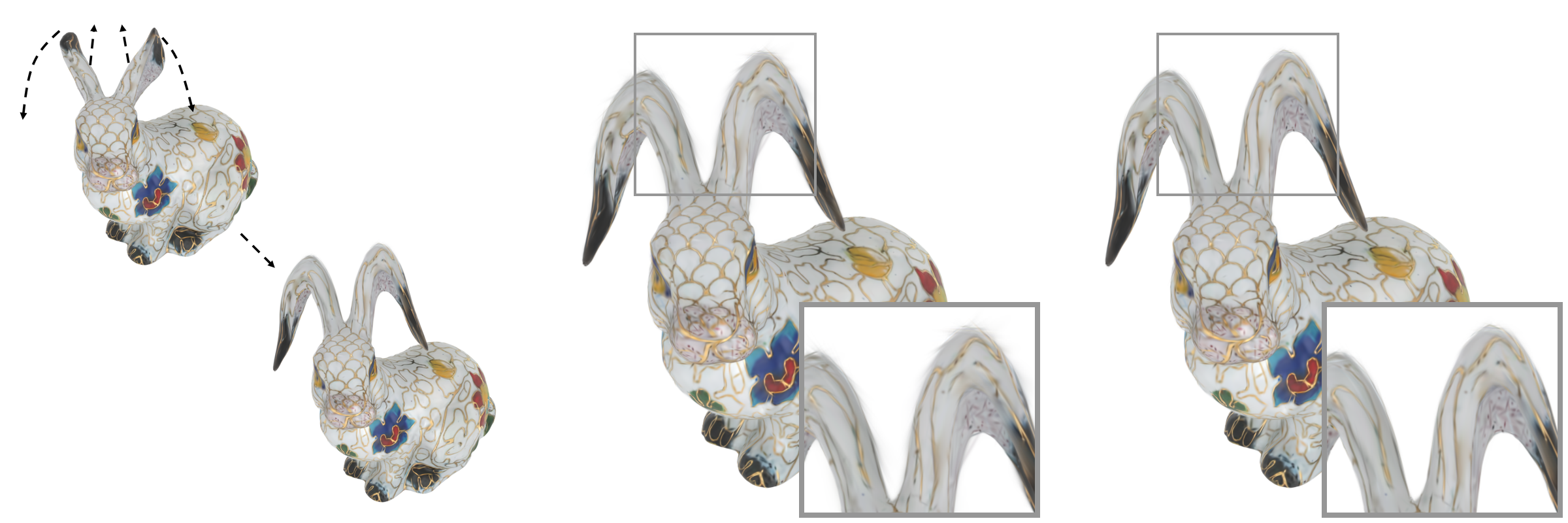}
    \end{subfigure}

    \begin{subfigure}[b]{0.49\textwidth}
        \centering
        (c) Pinocchio
    \end{subfigure}
    \hfill
    \begin{subfigure}[b]{0.49\textwidth}
        \centering
        (d) Bunny
    \end{subfigure}

    \begin{subfigure}[b]{0.49\textwidth}
        \centering
        \includegraphics[width=\textwidth]{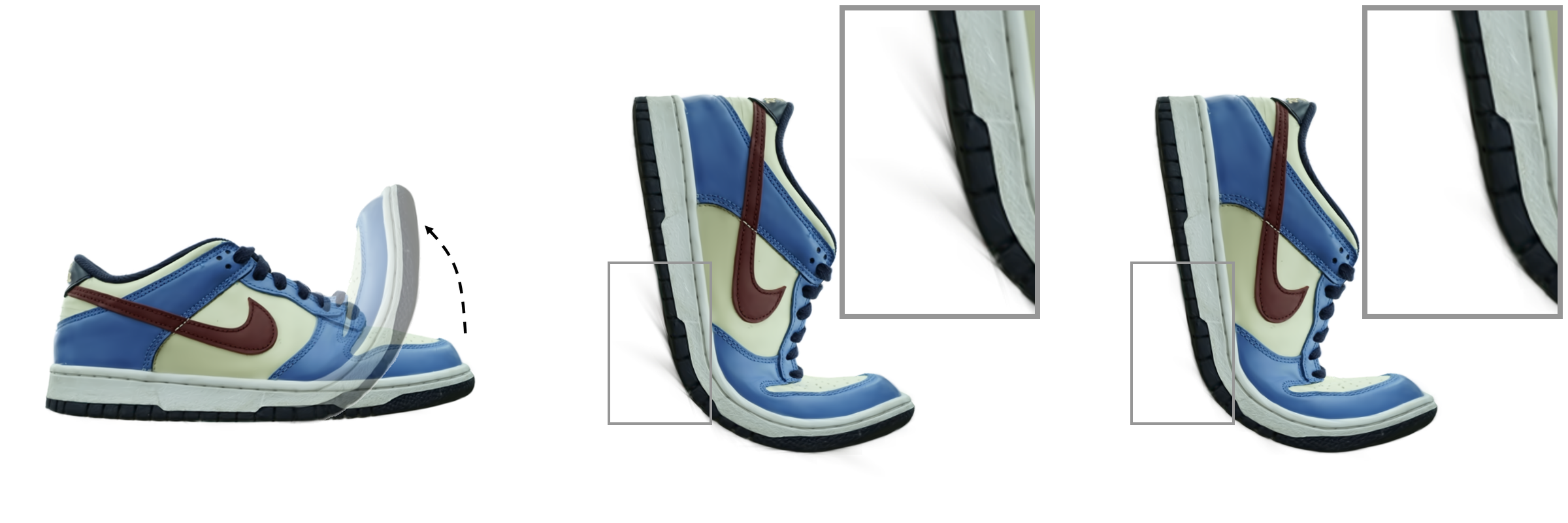}  
    \end{subfigure}
    \hfill
    \begin{subfigure}[b]{0.49\textwidth}
        \centering
        \includegraphics[width=\textwidth]{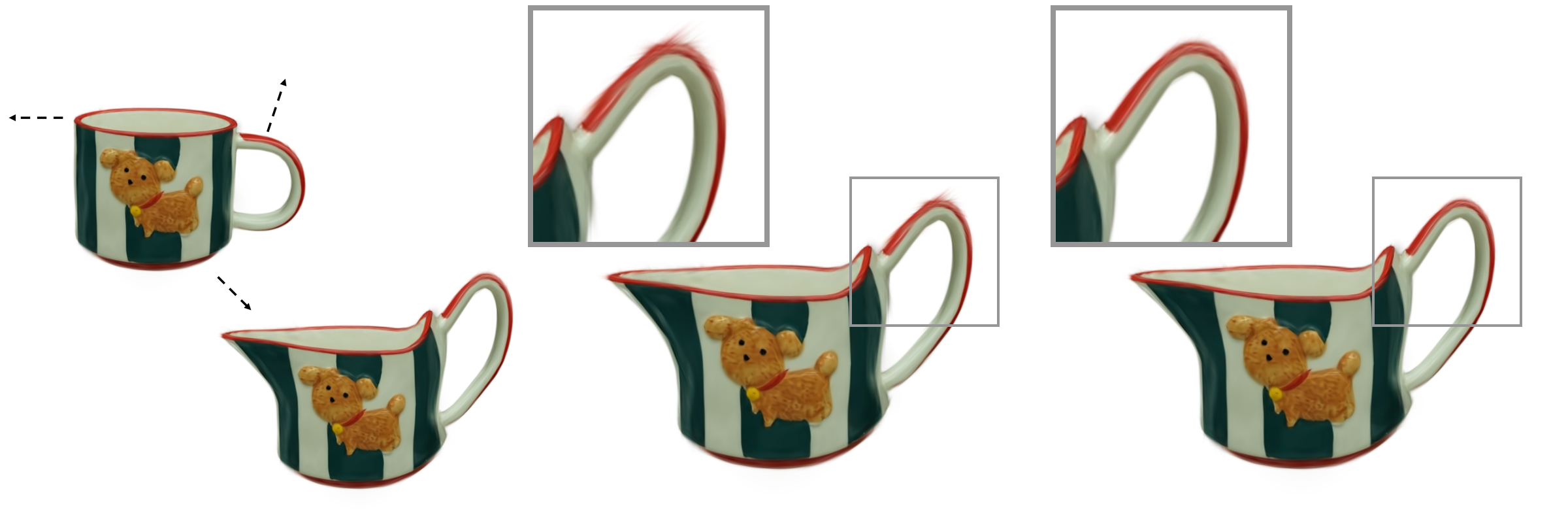}  
    \end{subfigure}

    \begin{subfigure}[b]{0.49\textwidth}
        \centering
        (e) Shoe
    \end{subfigure}
    \hfill
    \begin{subfigure}[b]{0.49\textwidth}
        \centering
        (f) Cup
    \end{subfigure}

    \caption{\textbf{Comparisons with the baseline.} The first two rows in the figure are synthetic data, while the third row shows the real-world data. The baseline refers to the results after only applying geometric deformation in Section~\ref{sec:stage1}. From the comparison, it is evident that our method outperforms the baseline under various deformations.}
    \label{fig:vsBaseline}
\end{figure*}

\subsubsection{Radial Features}
\label{sec:feature}
A 3DGS approximates a continuous radiance field, which can be regarded as the superposition of local fields associated with each Gaussian. To mathematically characterize this radiance field at an arbitrary position $\bm{p}$, we define two radial features namely $\bm{F}$ and $O$:
\begin{equation}\label{eq:feature}
    \bm{F}(\bm{p}) = \sum^n_{i = 1} \alpha_i G_i(\bm{p} - \bm{p}_i)\cdot \bm{H}_i,
\end{equation} 
and
\begin{equation}\label{eq:opacity}
   O(\bm{p}) = \sum_{i=1}^{n} \alpha_{i} G_{i}(\bm{p}-\bm{p}_{i}). 
\end{equation}
Here, $G_i(\bm{x}) = e^{-0.5\bm{x}^\top \bm{\Sigma}_i^{-1} \bm{x}}$ is the $i$-th Gaussian characterized by its position $\bm{p}_{i}$, covariance matrix $\bm{\Sigma}_{i}$ (computed from its rotation $\bm{r}_{i}$ and scaling $\bm{s}_{i}$), opacity $\alpha_{i}$, and spherical harmonics $\bm{H}_{i}$. $n$ gives the total number of Gaussians in the object. Intuitively, feature $\bm{F}$ encodes rotated harmonics color, and $O$ indicates the transparency at $\bm{p}$.


\subsubsection{Field-wise Feature Error}
\label{sec:energy}
Embedded deformation warps the original radiance field. It is expected that the deformed Gaussians capture and replicate the desired radiance field with their local transformations i.e., Eq.~\eqref{eq:local_transformation}. Unfortunately, per-Gaussian transformation only inherits the ARAP deformation from the nearby control points of the deformation graph, and it is not explicitly linked to the radiance field --- the latter determines the final rendering quality. This inconsistency induces rendering artifacts, especially under large and local deformations. In order to resolve this issue and improve the image quality of deformed 3DGS, we finetune Gaussian deformation by enforcing deformed 3DGS to have radial features i.e., Eqs.~\eqref{eq:feature}~and~\eqref{eq:opacity} matching the ground truth values. Specifically, we aim to suppress two errors:
\begin{equation}\label{energy_feature}
    E_{feature} = \int |\bm{F}^*(\bm{x}) - \bm{F}(\bm{x})| d \bm{x},
\end{equation}
and
\begin{equation}\label{energy_opacity}
    E_{opacity} = \int |O^*(\bm{x}) - O(\bm{x})| d \bm{x}.
\end{equation}
Here $\bm{F}^*$ and $O^*$ refer to the desired radial features of spherical harmonics and opacity. The integral of the features' difference over the entire field suggests the 
discrepancy between the new radiance field, which is warped by the deformation graph, and the actual field encoded with the deformed Gaussians. 

The difficulty lies in the fact that the above integral can hardly be evaluated analytically. To this end, we sample the radiance field numerically at $m$ sample points. The harmonic features and opacity features at those samples for the rest-shape radiance field are pre-computed. After applying the embedded deformation, we can conveniently obtain their new positions in the warped field. The harmonic features can also be computed based on the deformation gradient tensor at sample points. This information is our best bet for the ground truth radial features after the deformation, denoted with $\bm{F}^*$ and $O^*$ in our second-stage optimization. The radial features from deformation Gaussians, on the other hand, are computed in the standard way using Eqs.~\eqref{eq:feature}~and~\eqref{eq:opacity}. The numerical errors now become:
\begin{equation}\label{eq:energy_feature}
    E_{feature} \approx \sum_{j=1}^m |\bm{F}^*(\bm{x}_j) - \bm{F}(\bm{x}_j)|,
\end{equation}
and
\begin{equation}\label{energy_opacity}
    E_{opacity} \approx \sum_{j=1}^m |O^*(\bm{x}_j) - O(\bm{x}_j)|,
\end{equation}
as the summation of feature differences at all $m$ sample points $\bm{x}_j$.

Another challenging aspect of deforming 3DGS is at the boundary of an object i.e., Fig. \ref{fig:ab_boundary}. This is because the deformation graph is agnostic to the models represented by the radiance field. At the same time, it is often the case that users expect the deformation only to apply to the model rather than to the unoccupied empty space. In other words, we would like to ensure Embedded Deformation only triggers the deformation of the 3D model of interest, not the entire space, so that locally curving deformation can be effectively captured. This preference can be mathematically expressed as:
\begin{equation}\label{eq:energy_boundary}
   E_{boundary} = \sum_{\bm{x}_j \in \mathcal{O}} |O^*(\bm{x}_j) - O(\bm{x}_j)|,
\end{equation}
where $\mathcal{O} = \{\forall \bm{x}_j :  O^*(\bm{x}_j) = 0 \}$ denotes the set of sample points, which are likely to be at an unoccupied space. Putting together, we have the final error to be minimized:
\begin{equation}\label{eq:total_error}
    E_{total} = E_{feature} + \lambda_1 E_{opacity} + \lambda_2 E_{boundary}.
\end{equation}
$\lambda_1$ and $\lambda_2$ are two hyperparameters to balance different error types.

\noindent\textbf{Sampling \& Optimization}~~
We employ a simple and effective sampling scheme to measure $E_{total}$. At the rest shape, we extract an AABB for each Gaussian, and uniformly place sample points within bounding boxes. The adjacency to the control points on the deformation graph of those sample points is constructed with KNN. At each iteration of finding the optimal per-Gaussian deformation, the geometry of a Gaussian is changed, as well as their spatial relations with sample points (i.e., Eq.~\eqref{eq:feature}). We use spatial hashing to fast update this information. 

We employ the Adam optimizer~\cite{KingmaB14} to simultaneously optimize the position $\bm{p}$, rotation $\bm{r}$, scaling $\bm{s}$, spherical harmonics $\bm{H}$, and opacity $\alpha$ of the Gaussians using a gradient-based backpropagation. We utilize the sigmoid activation function for the opacity of Gaussians, as well as the exponential activation function for scaling, to enhance the stability of the optimization, as employed in the vanilla 3DGS~\cite{KerblKLD23}.

\begin{figure}[t]
    \centering
    \begin{subfigure}[b]{0.14\textwidth}
        \centering
        $\lambda_{lpf} = 0.02$
    \end{subfigure}
    \hfill
    \begin{subfigure}[b]{0.14\textwidth}
        \centering
        $\lambda_{lpf} = 0.2$
    \end{subfigure}
    \hfill
    \begin{subfigure}[b]{0.14\textwidth}
        \centering
        $\lambda_{lpf} = 2.0$
    \end{subfigure}

    \begin{subfigure}[b]{0.48\textwidth}
        \centering
        \includegraphics[width=\textwidth]{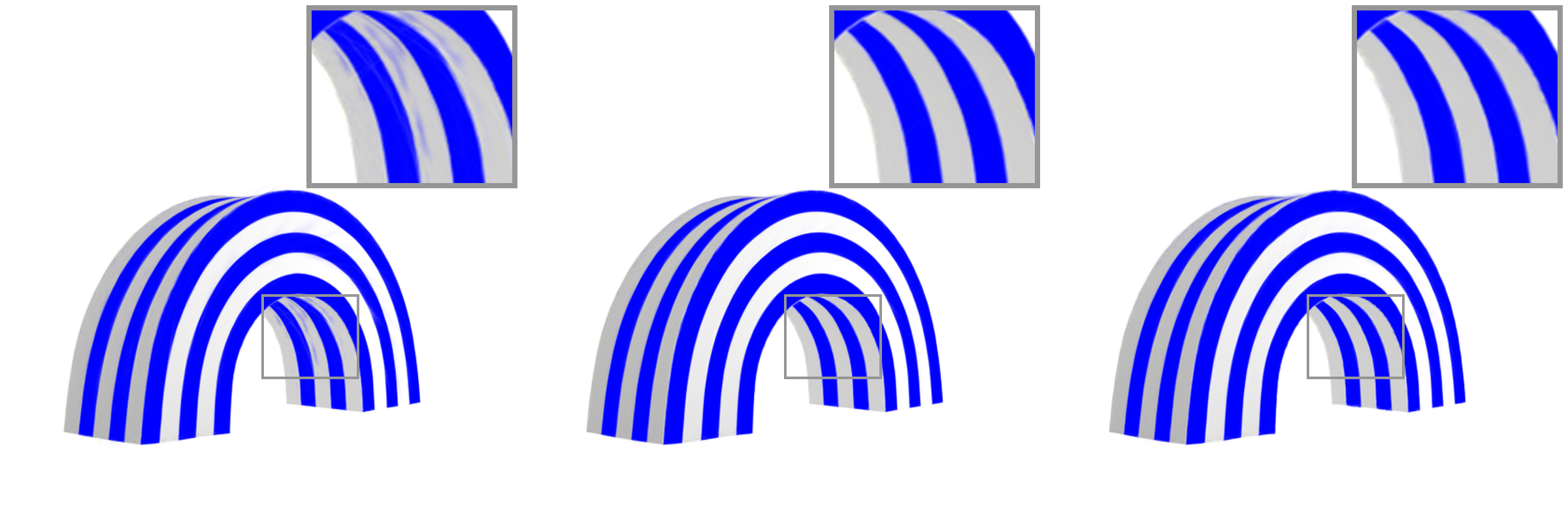}  
    \end{subfigure}
    \hfill
    \begin{subfigure}[b]{0.48\textwidth}
        \centering
        \includegraphics[width=\textwidth]{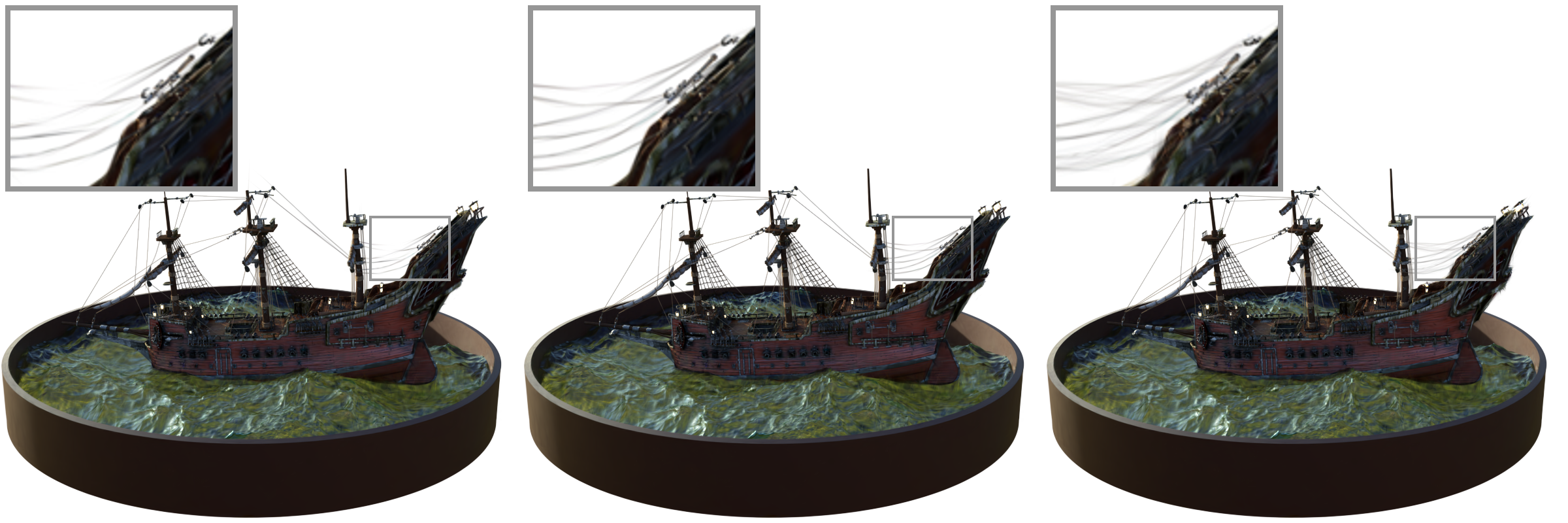}  
    \end{subfigure}
    \caption{\textbf{Different values of $\lambda_{lpf}$.} $\lambda_{lpf}$ controls the passing bandwidth of the low-pass filter. An over-conservative $\lambda_{lpf}$ (i.e., $\lambda_{lpf} = 0.02$) could lead to under-sampling. An over-aggressive $\lambda_{lpf}$, on the other hand, could bring blurriness. Nevertheless, the final results are not sensitive to $\lambda_{lpf}$, and we found setting $\lambda_{lpf}$ to $0.2$ yields good results in general.}
    \label{fig:lambda_lpf}
\end{figure}

\begin{figure*}[ht!]
    \centering
    \begin{subfigure}[b]{0.19\textwidth}
        \centering
        Input \& Deform
    \end{subfigure}
    \hfill
    \begin{subfigure}[b]{0.19\textwidth}
        \centering
        Mesh(GT)
    \end{subfigure}
    \hfill
    \begin{subfigure}[b]{0.19\textwidth}
        \centering
        GaussianMesh~\cite{Gao24}
    \end{subfigure}
    \hfill
    \begin{subfigure}[b]{0.19\textwidth}
        \centering
        Frosting~\cite{GSFrosting}
    \end{subfigure}
    \hfill
    \begin{subfigure}[b]{0.19\textwidth}
        \centering
        Ours
    \end{subfigure}

    \begin{subfigure}[b]{0.99\textwidth}
        \centering
        \includegraphics[width=\textwidth]{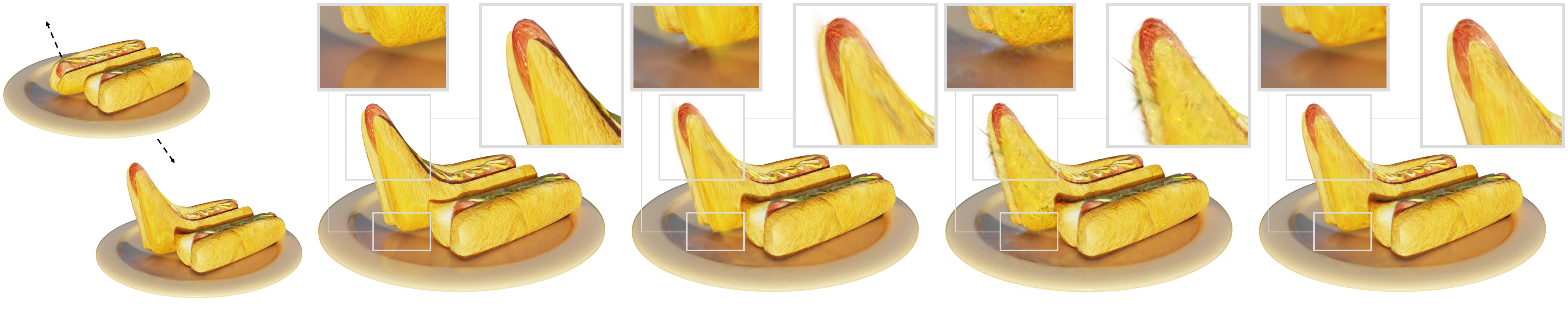}  
    \end{subfigure}
    \hfill
    \begin{subfigure}[b]{0.99\textwidth}
        \centering
        \includegraphics[width=\textwidth]{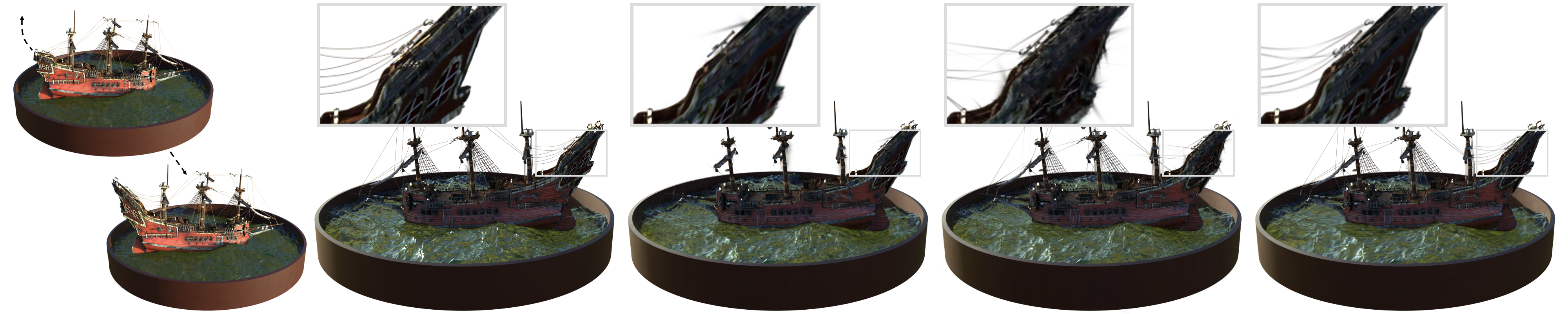}  
    \end{subfigure}
    \caption{\textbf{Comparisons with existing 3DGS deformation methods.} We present the deformed results of existing methods GaussianMesh~\cite{Gao24}, Gaussian Frosting~\cite{GSFrosting} and ours. The comparison demonstrates that our method better preserves Gaussian radiance fields after the same deformation.}
    \label{fig:vs_3DMethod}
\end{figure*}

\subsubsection{Adaptive Anisotropic Spatial Low-Pass Filter}
\label{sec:lpf}
The Nyquist-Shannon Sampling Theorem~\cite{Nyquist02, Shannon98} states that, to accurately reconstruct a band-limited signal, the discrete sampling frequency must be at least twice the highest frequency of the signal; otherwise, aliasing will occur. To address the aliasing problem caused by undersampling high-frequency Gaussians, EWA-Splatting~\cite{ZwickerPBG02} and 3DGS~\cite{KerblKLD23} employ a 2D low-pass filter on Gaussians in the projection plane, while Mip-Splatting~\cite{Mipsplatting} introduces a 3D low-pass filter for Gaussians in the spatial domain. 
Those strategies work well for a static scene. In our situation, the sampling intervals vary, and they are anisotropic during the deformation process. During the deformation process, if the distance among sampling points increases, indicating the area is being stretched, the size of the Gaussian provided by the low-pass filter should also be adjusted adaptively.

We estimate the post-deformation sampling intervals by calculating the $3 \times 3$ matrix $\bm{A}_k$ at each grid cell during the deformation. $\bm{A}_k$ offers the first-order approximation of the deformation field near the cell center, which can be computed in a similar way as Eq.~\eqref{eq:local_transformation}. $\bm{A}_k/n_{dim}$ serves a good estimation of the anisotropic sampling intervals within the grid cell, where $n_{dim}$ is the number of sample points per axis in each grid. We apply the following adaptive anisotropic low-pass filter: 
\begin{equation}
    \bm{\Sigma}_i = \bm{R}_i\bm{S}_i\bm{S}_i^\top\bm{R}_i^\top + \frac{\lambda_{lpf}}{n_{dim}^2}\bm{A}_k\bm{A}_k^\top
    \label{lpf}
\end{equation}
Here, $\bm{\Sigma}_i$, $\bm{R}_i$, $\bm{S}_i$ are respectively covariance matrix, rotation matrix, scaling matrix of $i$-th Gaussian, whose center is assumed located within the $k$-th grid cell. $\lambda_{lpf}$ is a hyperparameter that controls the low-pass filter. We empirically set it at $0.2$ in our experiments. The final results are not sensitive to $\lambda_{lpf}$ (Fig.~\ref{fig:lambda_lpf}).







\section{Experiments \& Evaluations}\label{sec:experiment}
We implemented our method on a single NVIDIA RTX 4090 24GB GPU, using C++ and CUDA for development. We have thoroughly validated our method in a wide range of 3DGS-based models. We note that our method outperforms both the baseline and existing methods in terms of visual quality and quantitative metrics. We also performed ablation studies on several core design elements to illustrate their individual contributions.  


\subsection{Datasets \& Metrics}
We conducted experiments on two real-world datasets and six synthetic datasets. Among those synthetic datasets, two were selected from the NeRF-Synthetic dataset, three models were from Sketchfab, and one was created by ourselves. For the synthetic datasets, we rendered and exported the corresponding camera poses using Blender. For the real-world datasets, we obtained the camera poses using COLMAP~\cite{SchonbergerF16, SchonbergerZFP16}.

In the synthetic dataset, we applied the same graph deformation to both our Gaussian radiance field and the vertices of the original mesh. The deformed mesh results were used as ground truth for computing quantitative metrics on the rendering results of the deformed Gaussian radiance field. For each synthetic dataset, We compare the Peak Signal-to-Noise Ratio (PSNR), Structural Similarity (SSIM)~\cite{WangBSS04}, and Learned Perceptual Image Patch Similarity (LPIPS)~\cite{ZhangIESW18} between 200 novel-views rendered results of Gaussians and the ground truth. The 200 novel views were sampled from a 360-degree circular camera trajectory, allowing observation of the deformation areas and the overall object. The rendered results of the test viewpoints are shown in the video.


\subsection{Comparisons \& Evaluations}
We compare our method with the baseline and existing methods GaussianMesh~\cite{Gao24}, Gaussian Frosting~\cite{GSFrosting}, and GaMeS~\cite{abs-2402-01459}, which are considered the latest work on Gaussians deformation or geometrical editing.

\subsubsection{Comparisons with the baseline}
We compared our method with the baseline both visually and quantitatively under the same deformation. In the visual comparison, Fig. \ref{fig:vsBaseline} demonstrates that our method outperforms the baseline, which only geometrically deforms Gaussians. The baseline results exhibit some artifacts such as spikes and blurring caused by deformation. Our method effectively mitigates these artifacts. The quantitative results of synthetic datasets presented in Table. \ref{tab:vsBaseline} also show that our method exceeds the baseline in terms of rendering metrics across all datasets.

\begin{table}[h]
    \caption{\textbf{Comparisons with the baseline on the synthetic datasets.} We conducted a quantitative comparison with the baseline on a synthetic dataset. The quantitative results indicate that our method consistently outperforms the baseline.}
    \centering
    \begin{tabular}{c|c|c|c|c} 
        \hline
        Dataset & Method & PSNR $\uparrow$ & SSIM $\uparrow$ & LPIPS $\downarrow$ \\
        \hline
        \multirow{2}{*}{Pinocchio} & Baseline & 36.32 & 0.9838 & 0.0390 \\
        & Ours & \textbf{41.67} & \textbf{0.9869} & \textbf{0.0354} \\
        \hline
        \multirow{2}{*}{Bunny} & Baseline & 32.74 & 0.9709 & 0.0370 \\
        & Ours & \textbf{33.82} & \textbf{0.9756} & \textbf{0.0310} \\
        \hline
        \multirow{2}{*}{Stripes} & Baseline & 31.90 & 0.9868 & 0.0141 \\
        & Ours & \textbf{32.89} & \textbf{0.9924} & \textbf{0.0051} \\
        \hline
        \multirow{2}{*}{Glove} & Baseline & 28.11 & 0.9780 & 0.0339 \\
        & Ours & \textbf{28.24} & \textbf{0.9791} & \textbf{0.0305} \\
        
        \hline
    \end{tabular}
    \label{tab:vsBaseline}
\end{table}

\subsubsection{Comparisons with Existing Methods}

We conducted comparative experiments on the GaussianMesh~\cite{Gao24}, Gaussian Frosting~\cite{GSFrosting}, GaMeS~\cite{abs-2402-01459}, applying each method, along with ours, to deform the reconstructed Gaussian objects. 

Both GaussianMesh~\cite{Gao24} and Gaussian Frosting~\cite{GSFrosting} bind 3D Gaussians to a mesh or a frosting layer in the reconstruction process, enabling corresponding edits of the Gaussians by driving the mesh. For a fair comparison, we provide the same input and spatial deformation to these two 3DGS deforming methods and ours. Specifically, we input synthetic images and camera poses for the reconstruction, and use the same graph embedded deformation to perform editing in each method. We adopt the parameters and workflows recommended by each project for reconstruction, mesh extraction and refinement, and deformation. Different from GaussianMesh~\cite{Gao24} and Gaussian Frosting~\cite{GSFrosting}, our approach doses not rely on the mesh and simply applies the vanilla 3DGS~\cite{KerblKLD23} reconstruction. Fig. \ref{fig:vs_3DMethod} demonstrates that our method outperforms other methods, as the rendered results from our method are closer to the deformed mesh, which serves as the ground truth. The reason is that other methods only perform geometric edits on the Gaussians without addressing the mismatch between the deformed Gaussians and the deformed radiance field. Additionally, these two mesh-dependent methods may lose fine-structured geometry during mesh extraction, which can lead to failures in either reconstruction or deformation. Gaussian Frosting~\cite{GSFrosting}, when editing Gaussians, utilizes a fast but less accurate matrix decomposition to estimate the scale and rotation of the Gaussians, allowing them to edit a large number of Gaussians in real-time and achieve good results in rigid transformations and skeleton-driven deformations. However, when the deformation matrix of the Gaussians deviates from pure rotation or pure scaling, their approximation results in significant errors, causing artifacts. Table. \ref{tab:vs_3DMethods} shows that our method surpasses other methods quantitatively.

\begin{figure}[t]
    \centering
    \begin{subfigure}[b]{0.16\textwidth}
        \centering
        Mesh(GT)
    \end{subfigure}
    \hfill
    \begin{subfigure}[b]{0.15\textwidth}
        \centering
        GaMeS~\cite{abs-2402-01459}
    \end{subfigure}
    \hfill
    \begin{subfigure}[b]{0.16\textwidth}
        \centering
        Ours (flat)
    \end{subfigure}

    \begin{subfigure}[b]{0.48\textwidth}
        \centering
        \includegraphics[width=\textwidth]{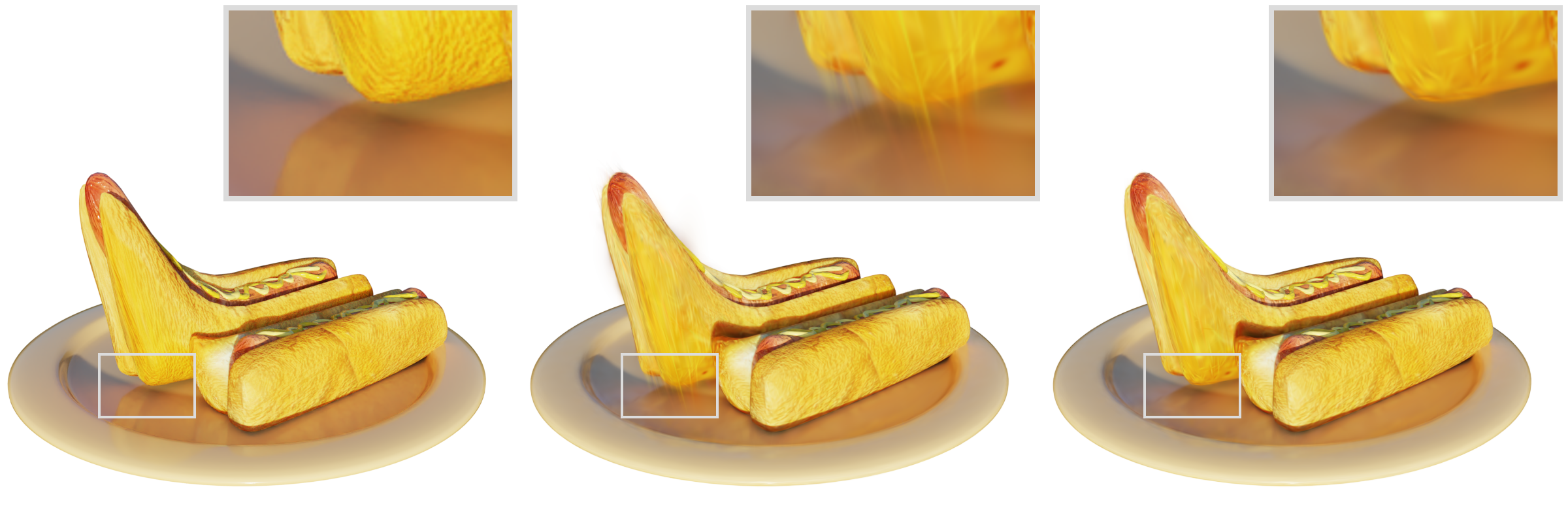}  
    \end{subfigure}
    \hfill
    \begin{subfigure}[b]{0.48\textwidth}
        \centering
        \includegraphics[width=\textwidth]{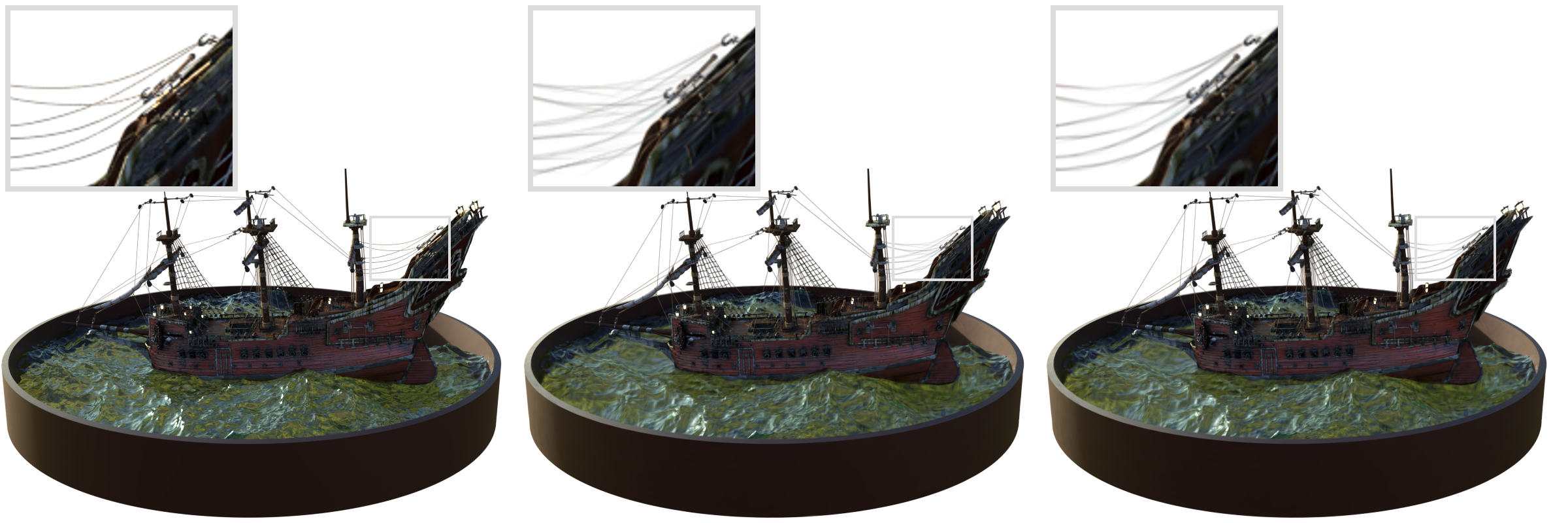}  
    \end{subfigure}
    \caption{\textbf{Comparisons with flat Gaussian editing methods.} We present the deformed results of flat Gaussian editing method, GaMeS~\cite{abs-2402-01459} and ours. The comparison shows that our method has better rendered results after performing the same deformation to the same flat Gaussian radiance field.}
    \label{fig:vsGaMeS}
\end{figure}

GaMeS~\cite{abs-2402-01459} offers a method for editing reconstructed flat Gaussians by manipulating the vertices of a triangle soup (pseudo-mesh). To facilitate a fair comparison of the deformation quality with the radiance fields represented by flat Gaussians in GaMeS, we utilize the reconstructed flat Gaussian object from GaMeS as the input for our deformations. Fig. \ref{fig:vsGaMeS} demonstrates that our method outperforms GaMeS. This arises from our consideration of the consistency of Gaussian radiance fields before and after deformation, while GaMeS focuses solely on geometric editing of flat Gaussians. The results also illustrates that our algorithm is effective on flat Gaussian objects. Table. \ref{tab:vsGaMeS} shows that our method quantitatively exceeds GaMeS.

Furthermore, all three existing Gaussian deformation methods require deformations to be performed within a specific Gaussian reconstruction pipeline or on a specified type of Gaussians. In contrast, our method takes any reconstructed Gaussians as input, without being constrained by the reconstruction process or the type of Gaussians used. 

\begin{table}[t]
    \caption{\textbf{Quantitative comparisons with existing 3DGS deformation methods on NeRF-synthetic dataset.} We performed quantitative comparisons with GaussianMesh~\cite{Gao24} and Gaussian Frosting~\cite{GSFrosting} on two datasets from the open-source NeRF-Synthetic Dataset. The quantitative results show that our method surpasses other methods on the deformed results with less number of Gaussians.}
    \centering
    \begin{tabular}{c|c|c|c|c|c} 
        \hline
        Dataset & Method & \# Gs $\downarrow$ & PSNR $\uparrow$ & SSIM $\uparrow$ & LPIPS $\downarrow$ \\
        \hline
        \multirow{3}{*}{Hotdog} & GsMesh~\cite{Gao24} & 453.3K & 27.87 & 0.9611 & 0.0371 \\
        & Frosting~\cite{GSFrosting} & 2.0M & 27.19 & 0.9465 & 0.0511 \\
        & Ours & \textbf{131.9K} & \textbf{28.02} & \textbf{0.9643} & \textbf{0.0326} \\
        \hline
        \multirow{3}{*}{Ship} & GsMesh~\cite{Gao24} & 621.2K & 26.21 & 0.9026 & 0.0889 \\
        & Frosting~\cite{GSFrosting} & 2.0M & 28.27 & 0.9077 & 0.0845 \\
        & Ours & \textbf{327.9K} & \textbf{29.59} & \textbf{0.9320} & \textbf{0.0578} \\
        \hline
    \end{tabular}
    \label{tab:vs_3DMethods}
\end{table}

\begin{table}[t]
    \caption{\textbf{Quantitative comparisons with flat Gaussian editing method on synthetic datasets.} We performed quantitative comparisons with GaMeS~\cite{abs-2402-01459}, a flat Gaussian editing method. The quantitative results show that our method outperforms GaMeS~\cite{abs-2402-01459} on the deformed results with flat Gaussian represented radiance fields.}
    \centering
    \begin{tabular}{c|c|c|c|c} 
        \hline
        Dataset & Method & PSNR $\uparrow$ & SSIM $\uparrow$ & LPIPS $\downarrow$ \\
        \hline
        \multirow{2}{*}{Hotdog} 
        & GaMeS~\cite{abs-2402-01459} & 28.35 & 0.9645 & 0.0337 \\
        & Ours (flat) & \textbf{28.41} & \textbf{0.9646} & \textbf{0.0324} \\
        \hline
        \multirow{2}{*}{Ship} 
        & GaMeS~\cite{abs-2402-01459} & 29.60 & 0.9318 & 0.0577 \\
        & Ours (flat) & \textbf{29.62} & \textbf{0.9322} & \textbf{0.0574} \\
        \hline
    \end{tabular}
    \label{tab:vsGaMeS}
\end{table}

\subsubsection{Performance Evaluation}
\label{sec:performance}
Table. \ref{tab:performance} presents the deformation and the optimization efficiency of our method. The optimization process converges within 100 iterations, with detailed convergence information provided in Section \ref{app:converges}. In all the cases demonstrated in this paper (except as specifically noted cases in the ablation study), our method was optimized 100 iterations. The deformation time (Def. Time) reported in the table represents the average time per deformation step, while the optimization time (Opt. Time) reflects the total duration of the optimization process (100 iterations). In conclusion, the FPS during deforming process is over 10, which is also related to the number of control nodes and free nodes in the deform graph. 

During the precomputation stage, the graph is constructed, the Gaussian radiance fields are sampled, and the k-nearest control points for the Gaussian endpoints and sampled points are determined. These one-time preprocessing overheads typically require only 1 to 3 seconds for most datasets. The details are provided in the supplementary material.

\begin{table}[t]
    \caption{\textbf{Deformation and optimization efficiency across datasets.} The table presents the time required to perform the deformations on the Gaussian objects shown in the paper, including the average time per deformation step and the total optimization time. The results demonstrate that our interactive system supports smooth deformation interactions, with users needing to wait within two seconds for optimization after the deformation process for most cases.}
    \centering
    \begin{tabular}{c|c|c|c|c} 
        \hline
        Dataset & \# Gaussians & \# Samples & Def. Time & Opt. Time \\
        \hline
        Pinocchio & 29.6K & 3.17M & 3ms & 1.01s \\
        \hline
        Bunny & 31.3K & 5.73M & 9ms & 1.05s \\
        \hline
        Stripes & 20.6K & 0.90M & 70ms & 1.92s \\
        \hline
        Glove & 16.1K & 1.13M & 13ms & 1.70s \\
        \hline
        Shoe & 25.8K & 2.10M & 17ms & 1.85s \\
        \hline
        Cup & 10.5K & 3.12M & 27ms & 2.23s \\
        \hline
        Hotdog & 131.9K & 2.50M & 23ms & 1.89s \\
        \hline
        Ship & 327.9K & 5.01M & 42ms & 3.42s \\
        \hline
    \end{tabular}
    \label{tab:performance}
\end{table}

\subsection{Ablation Study}
\subsubsection{Shape Energy}
\begin{figure}[t]
    \centering
    \begin{subfigure}[b]{0.14\textwidth}
        \centering
        Ours \\ (100 iters)
    \end{subfigure}
    \hfill
    \begin{subfigure}[b]{0.14\textwidth}
        \centering
        Ours \\ (1000 iters)
    \end{subfigure}
    \hfill
    \begin{subfigure}[b]{0.14\textwidth}
        \centering
        w/o ${E}_{shape}$ \\ (1000 iters)
    \end{subfigure}

    \begin{subfigure}[b]{0.48\textwidth}
        \centering
        \includegraphics[width=\textwidth]{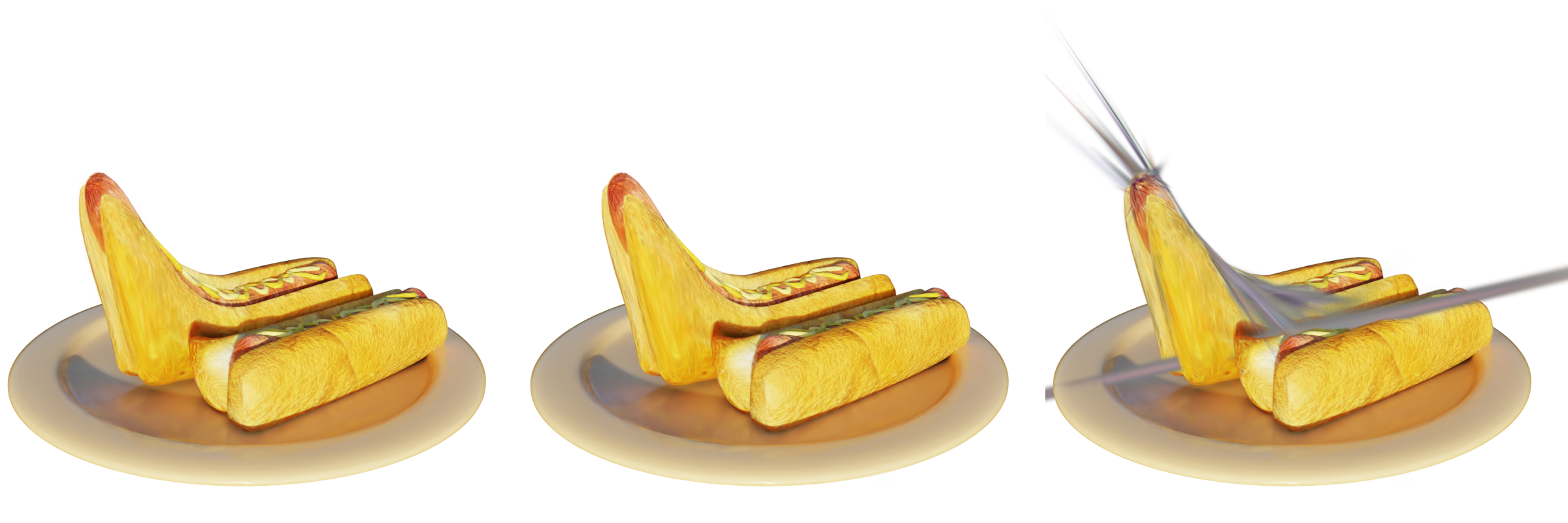}  
    \end{subfigure}
    \hfill
    \begin{subfigure}[b]{0.48\textwidth}
        \centering
        \includegraphics[width=\textwidth]{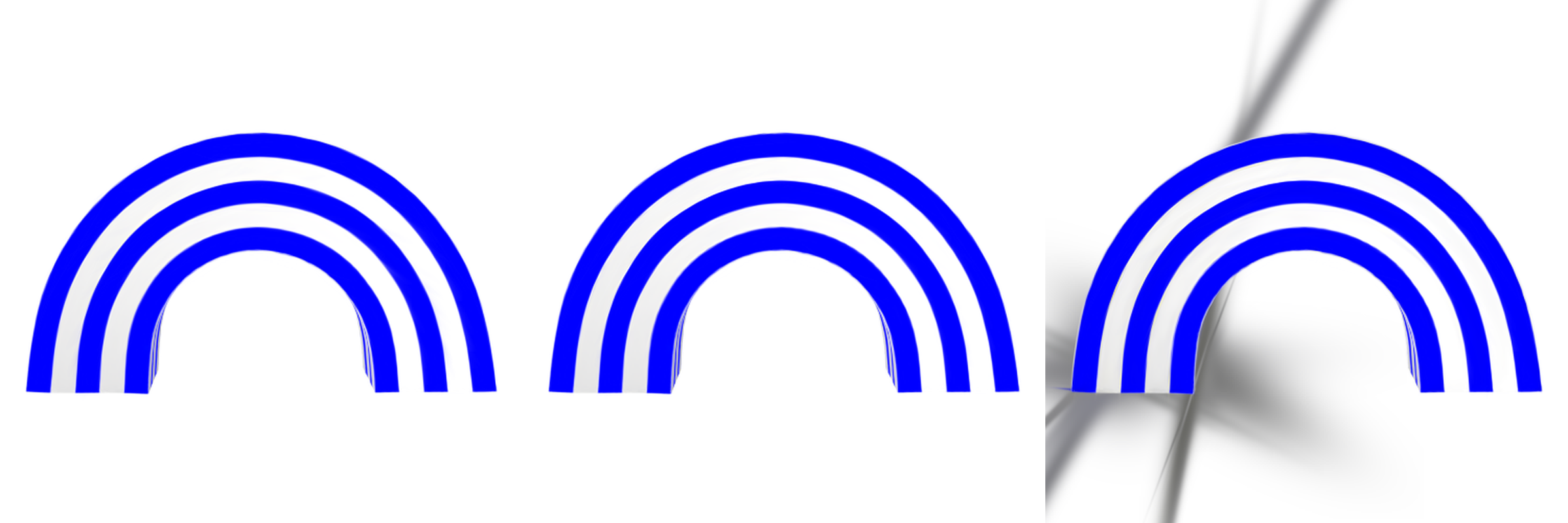}  
    \end{subfigure}
    \caption{\textbf{Ablation study on ${E}_{shape}$.} In the absence of the ${E}_{shape}$ constraint, the optimization process may become unstable. Some gray Gaussians, whose SH values are close to zero, cannot be adequately constrained with only the energy term ${E}_{feature}$. As shown in the figure, after introducing ${E}_{shape}$, our optimization process becomes stable, effectively resolving the aforementioned issue.}
    \label{fig:ab_shape}
\end{figure}

We define the energy that affects the shape of Gaussian radiance fields, including opacity energy and boundary energy, as shape energy ${E}_{shape}$. Ablation experiments were conducted on this shape energy. Relying solely on the feature energy term was insufficient to preserve the opacity and unitized colors (spherical harmonics) at each spatial sample point in the radiance field. We showcase the results of ablating the shape energy in Fig. \ref{fig:ab_shape}. The results demonstrate that without adding the shape energy term, it is impossible to sustain the shape and colors (spherical harmonics) of the Gaussian radiance fields. Some gray Gaussians may appear, with SH values close to zero, which cannot be constrained in the absence of ${E}_{shape}$. Consequently, the size of these gray Gaussians may become uncontrolled during the optimization process. With the inclusion of the shape energy term, these artifacts are effectively mitigated.

\subsubsection{Boundary Energy}
\begin{figure}[t]
    \centering
    \begin{subfigure}[b]{0.14\textwidth}
        \centering
        Baseline
    \end{subfigure}
    \hfill
    \begin{subfigure}[b]{0.14\textwidth}
        \centering
        w/o ${E}_{boundary}$
    \end{subfigure}
    \hfill
    \begin{subfigure}[b]{0.14\textwidth}
        \centering
        Ours
    \end{subfigure}

    \begin{subfigure}[b]{0.48\textwidth}
        \centering
        \includegraphics[width=\textwidth]{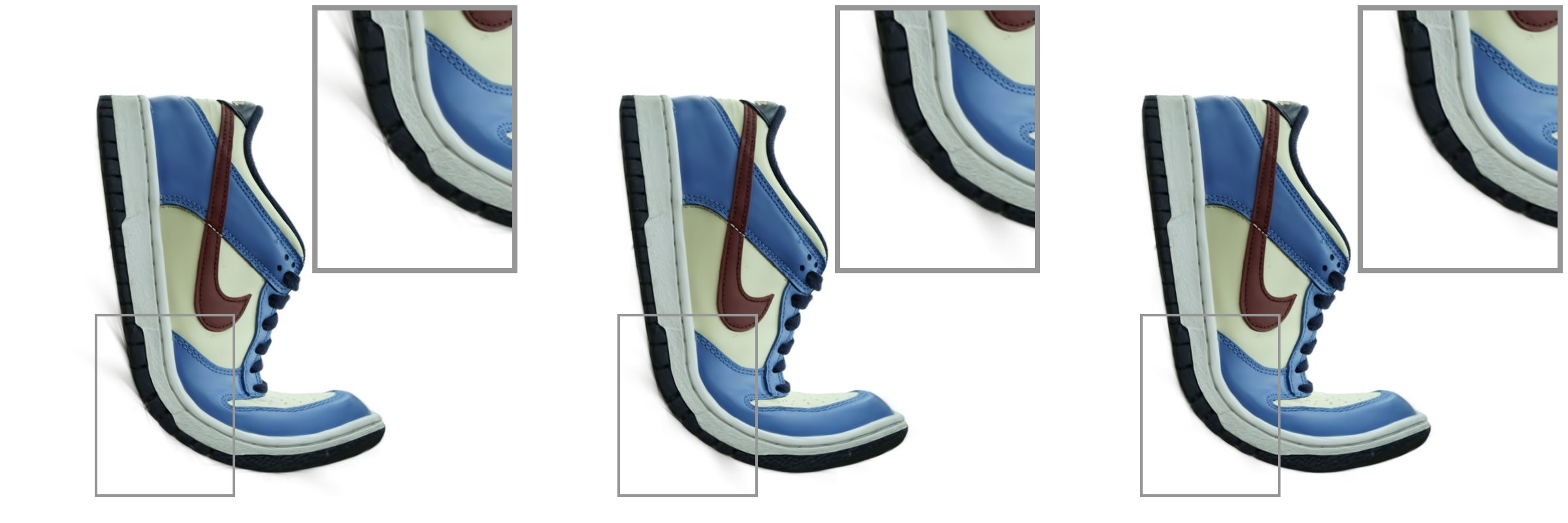}  
    \end{subfigure}

    \caption{\textbf{Ablation study on ${E}_{boundary}$.} As shown in the figure, our method better preserves the object's boundaries after deformation compared to the optimized results without the boundary energy term.}
    \label{fig:ab_boundary}
\end{figure}

When only feature energy and opacity energy are considered, some Gaussians with larger scales may still have small portions outside the object boundaries when optimization convergence. This occurs because the energy terms of these large Gaussians inside and outside the object are relatively balanced. Therefore, we introduced an additional boundary energy term to address such artifacts. The ablation study of the boundary energy is illustrated in Fig. \ref{fig:ab_boundary}. It can be observed that the boundary energy effectively resolves some artifacts protruding from the object boundaries.

\subsubsection{Adaptive Anisotropic Spatial Low-Pass Filter}
\begin{figure}[t]
    \centering
    \begin{subfigure}[b]{0.14\textwidth}
        \centering
        Baseline
    \end{subfigure}
    \hfill
    \begin{subfigure}[b]{0.14\textwidth}
        \centering
        w/o lpf
    \end{subfigure}
    \hfill
    \begin{subfigure}[b]{0.14\textwidth}
        \centering
        Ours
    \end{subfigure}

    \begin{subfigure}[b]{0.48\textwidth}
        \centering
        \includegraphics[width=\textwidth]{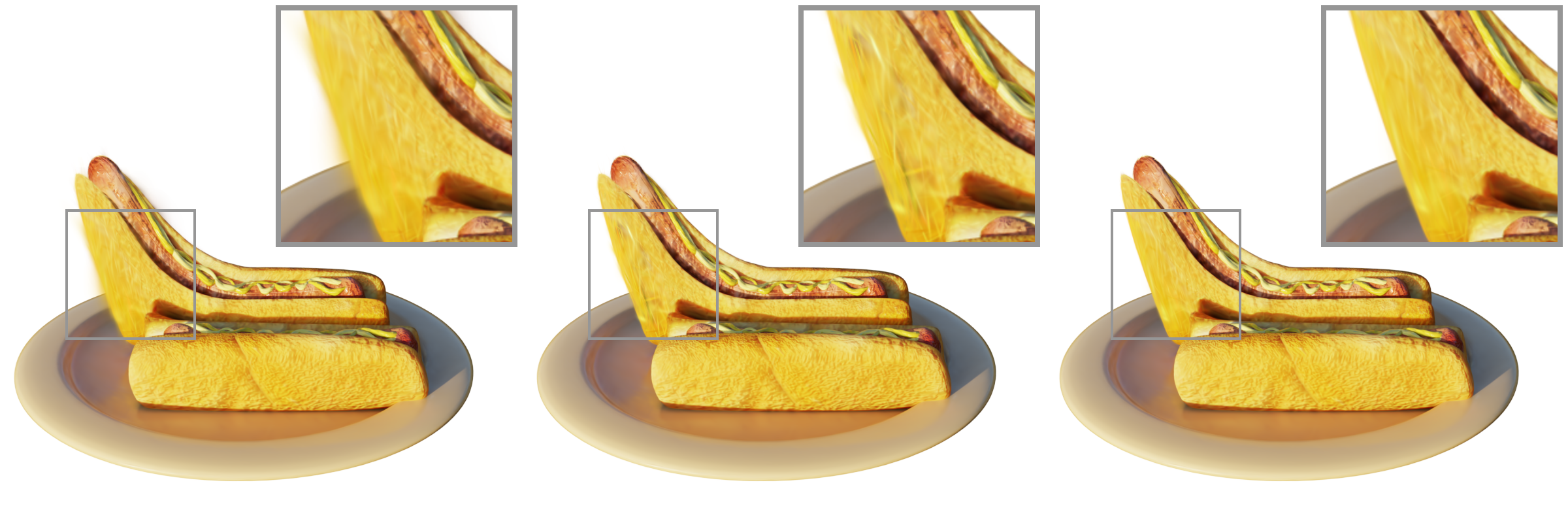}  
    \end{subfigure}
    \hfill
    \begin{subfigure}[b]{0.48\textwidth}
        \centering
        \includegraphics[width=\textwidth]{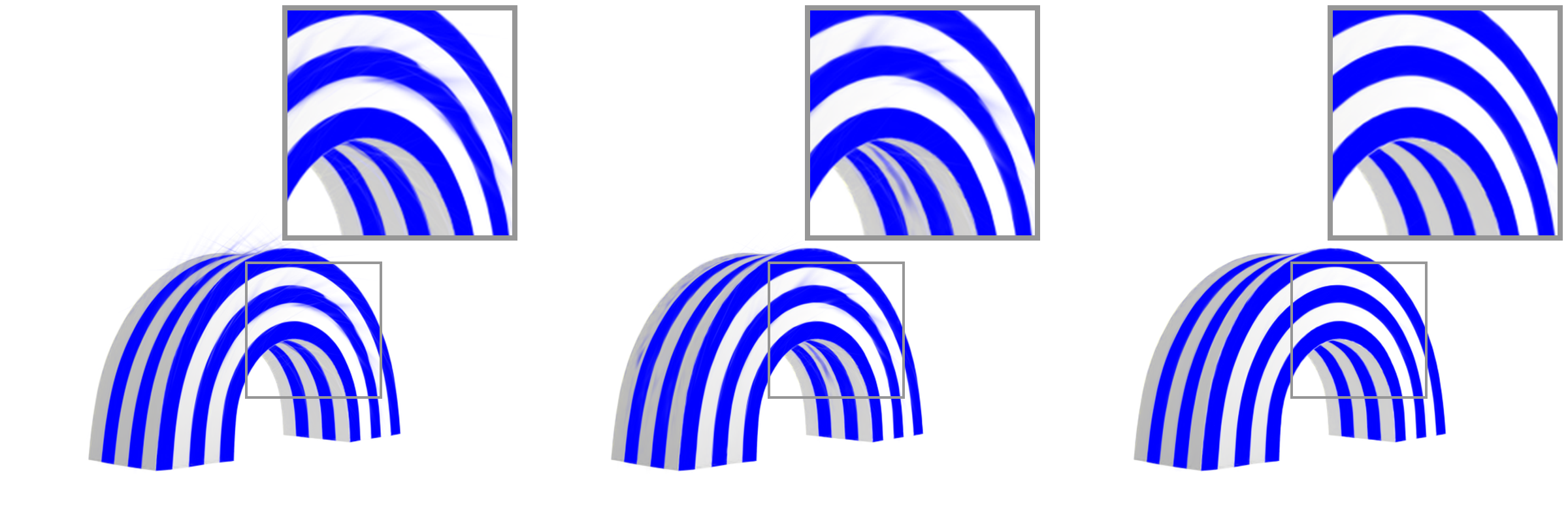}  
    \end{subfigure}
    
    \caption{\textbf{Ablation study on using spatial low-pass filter.} The comparison in the figure shows that when a spatial low-pass filter is not applied, the optimization results may exhibit artifacts such as holes due to undersampling of the original radiance field's spatial signals. In some regions, the results without low-pass filter are even worse than the baseline. This issue is resolved after applying our designed spatial low-pass filter.}
    \label{fig:ab_no_lpf}
\end{figure}

\begin{figure}[h!]
    \centering
    \begin{subfigure}[b]{0.14\textwidth}
        \centering
        Mesh(GT)
    \end{subfigure}
    \hfill
    \begin{subfigure}[b]{0.14\textwidth}
        \centering
        Fixed lpf
    \end{subfigure}
    \hfill
    \begin{subfigure}[b]{0.14\textwidth}
        \centering
        Ours
    \end{subfigure}

    \begin{subfigure}[b]{0.48\textwidth}
        \centering
        \includegraphics[width=\textwidth]{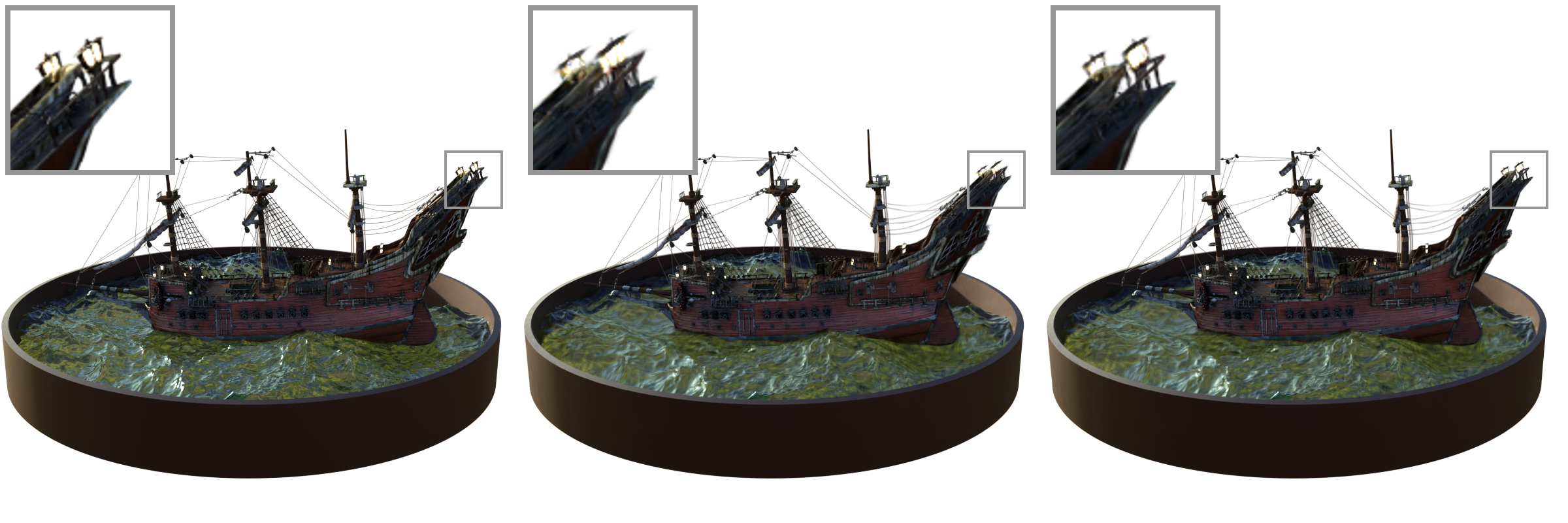}  
    \end{subfigure}
    
    \caption{\textbf{Ablation study on adaptive anisotropic spatial low-pass filter.} In cases where deformation leads to changes in sampling intervals and non-uniform distribution of sample points in space, using a fixed isotropic low-pass filter can result in some blurring. In contrast, our designed adaptive anisotropic spatial low-pass filter better preserves high-frequency signals in the radiance field when sampling intervals change.}
    \label{fig:ab_lpf}
\end{figure}

Since the spatial radiance field expressed by the Gaussians may contain high-frequency features, applying a low-pass filter to each Gaussian is a common approach to capture these high-frequency signals. During deformation, the interval of spatial sampling also changes. To address these variations in sampling intervals, we designed an adaptive anisotropic low-pass filter. To demonstrate the effectiveness of the adaptive anisotropic low-pass filter, we compared the results of not using a low-pass filter, using a fixed isotropic low-pass filter, and using the adaptive anisotropic low-pass filter, as shown in Fig. \ref{fig:ab_no_lpf} and \ref{fig:ab_lpf}. When a spatial low-pass filter is not applied, the optimization results may exhibit holes. Using a fixed isotropic low-pass filter can cause blurring in regions where the samples' intervals changes due to deformation. The comparison reveals that our design reduces the occurrence of artifacts and better preserves the original spatial signals.

\subsection{Selection of Parameters and Optimization Steps}

\subsubsection{K-value Selection in KNN}
\label{app:knn}

\begin{figure}[t]
    \centering
    \begin{subfigure}[b]{0.14\textwidth}
        \centering
        4NN
    \end{subfigure}
    \hfill
    \begin{subfigure}[b]{0.14\textwidth}
        \centering
        7NN
    \end{subfigure}
    \hfill
    \begin{subfigure}[b]{0.14\textwidth}
        \centering
        10NN
    \end{subfigure}

    \begin{subfigure}[b]{0.48\textwidth}
        \centering
        \includegraphics[width=\textwidth]{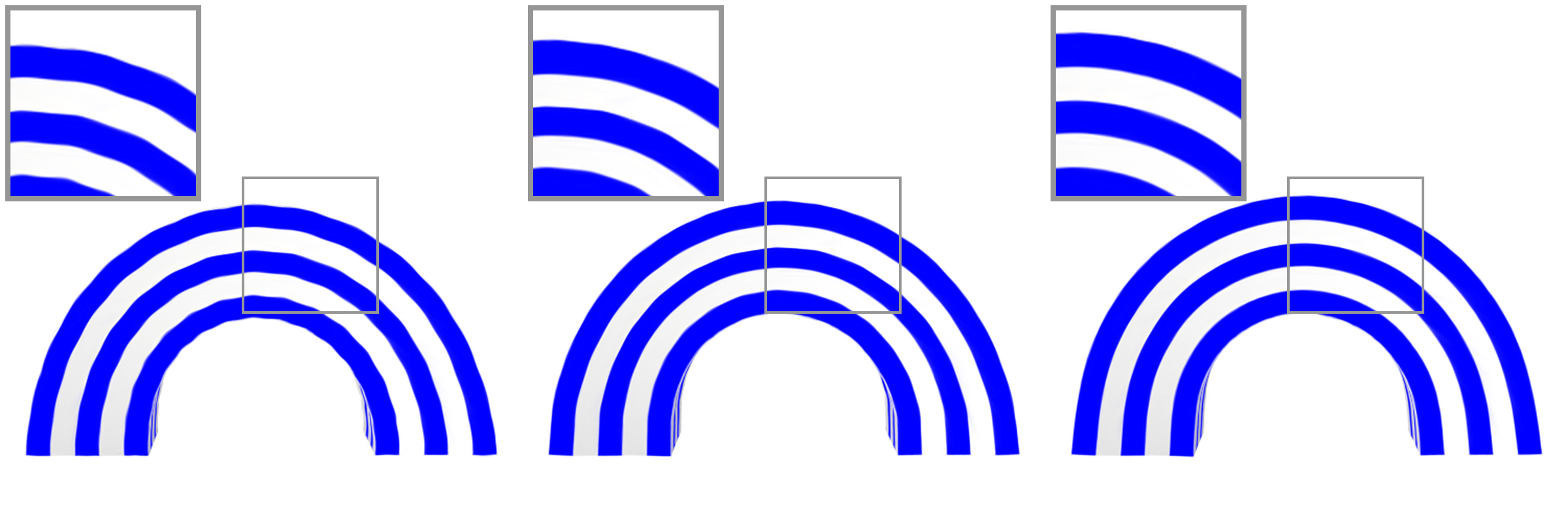}  
    \end{subfigure}
    
    \caption{\textbf{The experiments on different values of k in KNN.} In the three examples shown in the figure, we used the same Gaussian object and applied the same edits to the positions of the control points. It can be observed that larger k-values in KNN produce smoother results.}
    \label{fig:knn}
\end{figure}

We conducted experiments using KNN with different k-values. The deformation and optimization results from our algorithm are shown in the Fig. \ref{fig:knn}. For some regular geometries, it is evident that larger k-values result in smoother deformations of the Gaussian radiance fields. This finding is consistent with the experimental results on meshes in Embedded Deformation~\cite{SumnerSP07}.

\subsubsection{Optimization Convergence and Stability}
\label{app:converges}

We conducted quantitative experiments to demonstrate that our optimization process yields stable results, and that the optimization has converged to a good result after 100 iterations. For all six synthetic datasets in the paper, we applied the deformation presented earlier and performed a quantitative analysis by comparing the results without optimization, after 100 iterations of optimization, and after 1000 iterations of optimization with the ground truth. The average quantitative results on these datasets are presented in Table. \ref{tab:convergency}. It demonstrates that the energy terms and the optimization approach we designed for the Gaussians produce stable rendering results from novel views. Additionally, 100 iterations of optimization are sufficient to achieve a good and stable outcome.

\begin{figure*}[t]
    \centering
    \begin{minipage}[c]{0.02\textwidth}
        \rotatebox{90}{Output \hspace{1.8cm} Input \hspace{1.8cm} Output \hspace{1.8cm} Input \hspace{0.5cm}}
    \end{minipage}%
    \begin{minipage}[c]{0.98\textwidth}
        \begin{subfigure}[b]{0.16\textwidth}
            \centering
            View 0
        \end{subfigure}
        \hfill
        \begin{subfigure}[b]{0.16\textwidth}
            \centering
            View 1
        \end{subfigure}
        \hfill
        \begin{subfigure}[b]{0.16\textwidth}
            \centering
            View 2
        \end{subfigure}
        \begin{subfigure}[b]{0.16\textwidth}
            \centering
            View 0
        \end{subfigure}
        \hfill
        \begin{subfigure}[b]{0.16\textwidth}
            \centering
            View 1
        \end{subfigure}
        \hfill
        \begin{subfigure}[b]{0.16\textwidth}
            \centering
            View 2
        \end{subfigure}

        \begin{subfigure}[b]{0.49\textwidth}
            \centering
            \includegraphics[width=\textwidth]{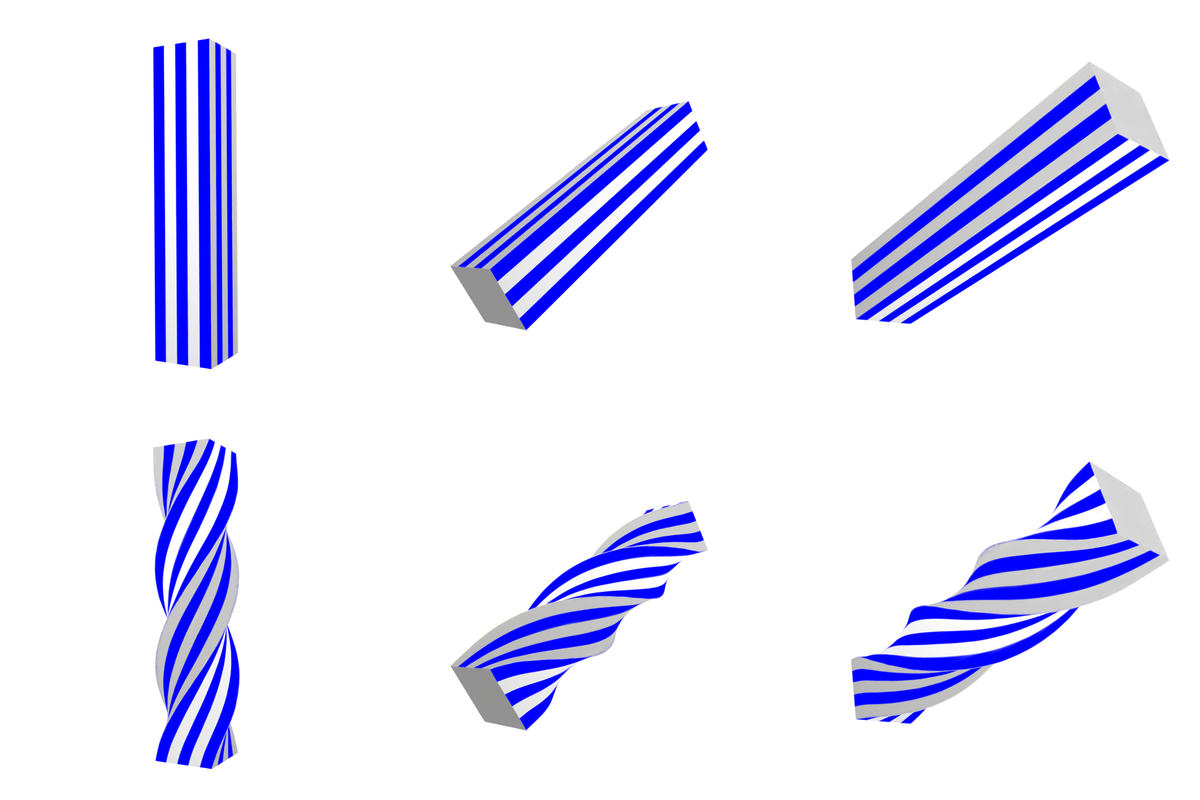}  
        \end{subfigure}
        \hfill
        \begin{subfigure}[b]{0.49\textwidth}
            \centering
            \includegraphics[width=\textwidth]{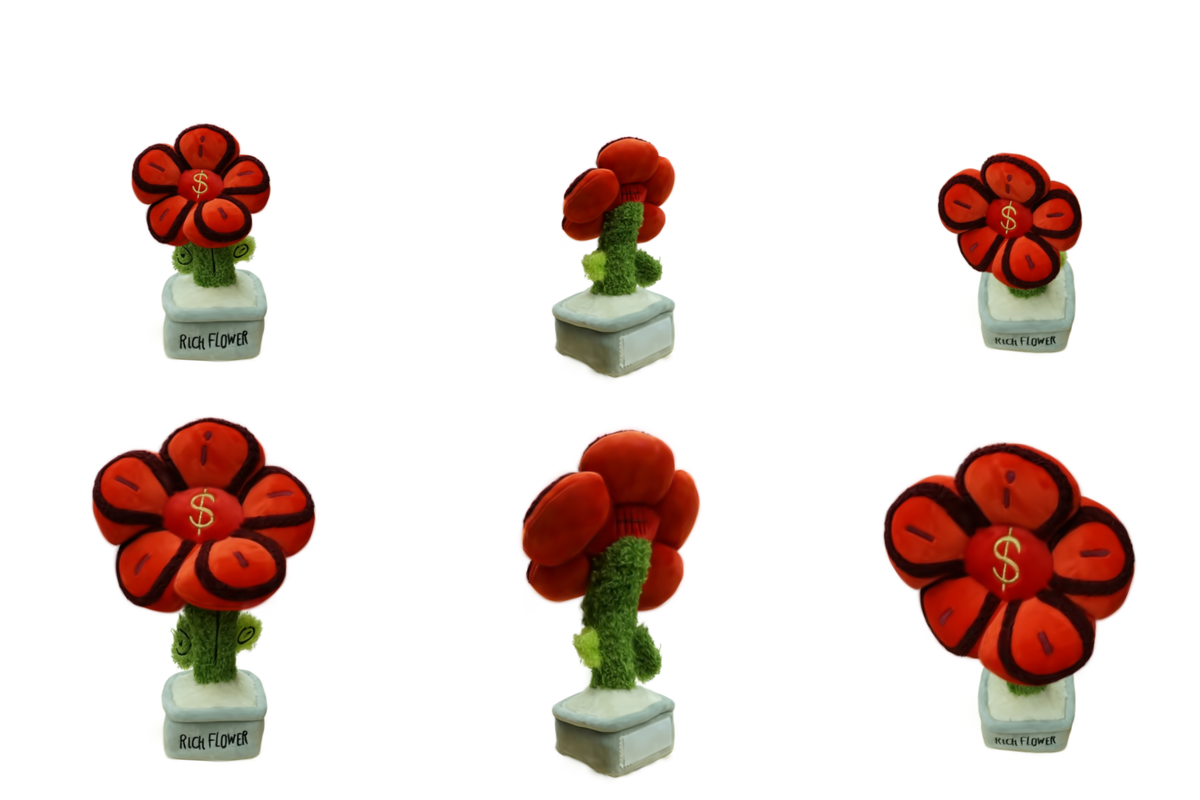}  
        \end{subfigure}
    
        \begin{subfigure}[b]{0.49\textwidth}
            \centering
            \includegraphics[width=\textwidth]{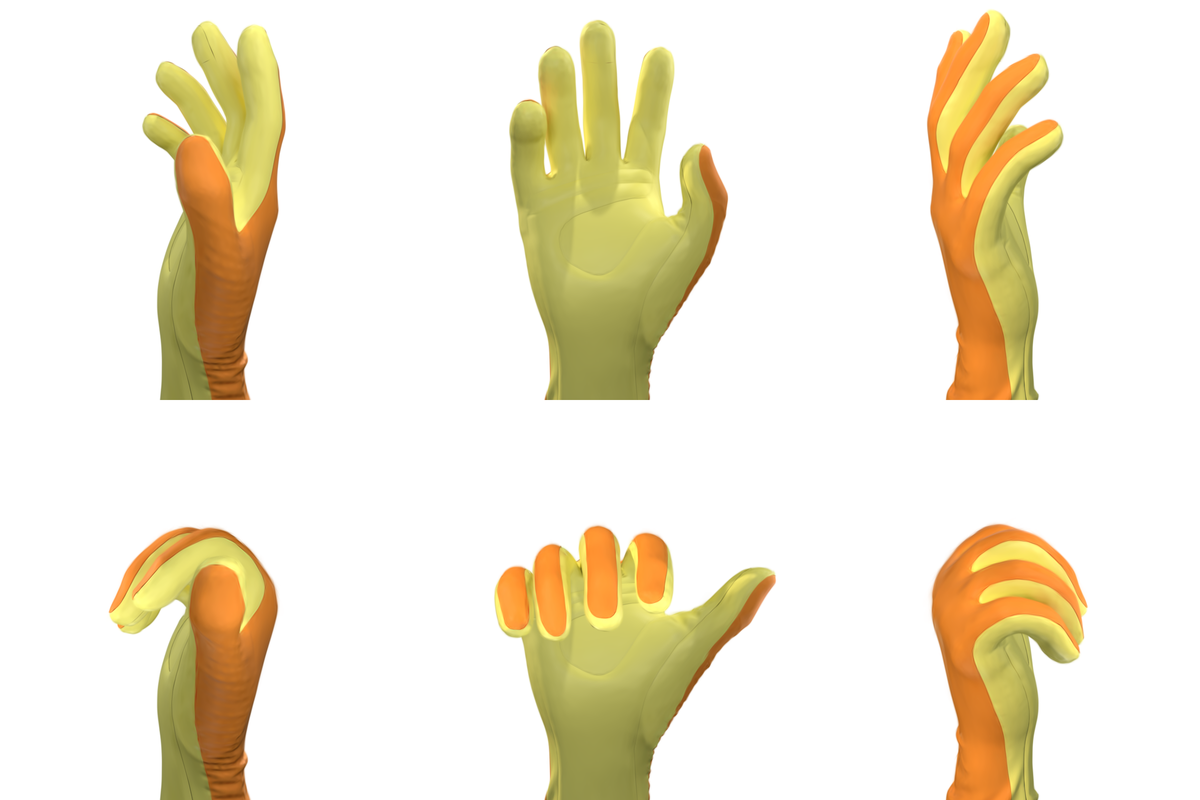}  
        \end{subfigure}
        \hfill
        \begin{subfigure}[b]{0.49\textwidth}
            \centering
            \includegraphics[width=\textwidth]{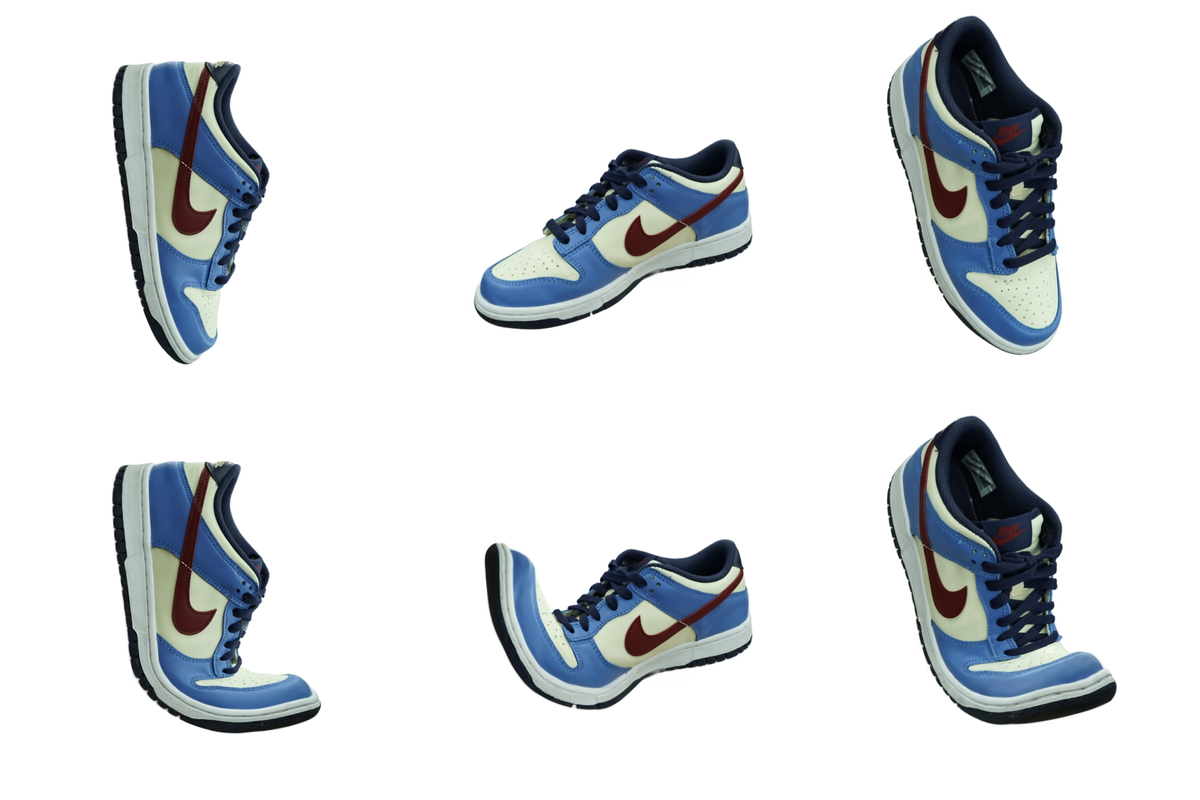}  
        \end{subfigure}
    \end{minipage}
    \caption{\textbf{Additional results of various types of deformations in novel-view synthesis.} In the figure, we illustrate deformations such as twisting, scaling, and bending. The flower dataset is a real-world dataset.}
    \label{fig:moreresults}
\end{figure*}

\subsection{Implementation Details}\label{sec:details}
During the initialization of sample points, we mark a grid cell as active if it or its neighbor contains non-zero feature values. We only update features at sample points in those active grid cells after the deformation to avoid unneeded computations.
During the optimization process, we do not update the properties of non-moving Gaussians. This prevents the optimization region from spreading from local deformations to the entire domain, avoiding unnecessary computational overhead.


We use $\lambda_1 = 1$, $\lambda_2 = 99$ in all experiments. We use $E_{feature}$ to preserve the accumulated color and $E_{opacity}$ to preserve the accumulated opacity in the desired radiance field. We think these two terms contribute equally to the rendering quality, and assign them equal weights. Additionally, we employ $E_{boundary}$ to ensure that there are no non-transparent Gaussians in the empty regions of the desired radiance field. Since the scope of $E_{boundary}$ differs from the $E_{feature}$ and $E_{opacity}$, we assign it a large weight to guarantee its dominance in empty regions. The hyperparameter $\lambda_{lpf} = 0.2$ that we use in the adaptive anisotropic 3D low-pass filter is the same as the parameter used in the spatial low-pass filter in Mip-Splatting~\cite{Mipsplatting}. For the datasets used in the experiments, we set the value of $k$ in KNN to 8, except for the stripes dataset, where $k$ was set to 10 to achieve smoother deformation results. The experiment analyzing the impact of the $k$ value in KNN on deformation results is detailed in Section \ref{app:knn}. All experiments demonstrated in this paper have been tested and can be executed on a single GPU with 8GB of VRAM. 

\begin{table}[t]
    \caption{\textbf{The experiments on optimization convergence and stability.} The table presents the average quantitative results across all six synthetic datasets in the paper (based on 3DGS representation). The three rows correspond to the results without optimization, after 100 iterations of optimization, and after 1000 iterations of optimization, respectively.}
    \centering
    \begin{tabular}{c|c|c|c} 
        \hline
        Method & PSNR $\uparrow$ & SSIM $\uparrow$ & LPIPS $\downarrow$ \\
        \hline
        Baseline & 31.06 & 0.9688 & 0.0363 \\
        Ours(100 iters) & 32.37 & 0.9717 & 0.0321 \\
        Ours(1000 iters) & 32.39 & 0.9718 & 0.0321 \\
        
        \hline
    \end{tabular}
    \label{tab:convergency}
\end{table}

During the optimization process, the learning rates for the rotation $\bm{r}$, scaling $\bm{s}$, opacity $\alpha$ and spherical harmonics $\bm{H}$ of Gaussians are set to $2 \times 10^{-3}$, $5 \times 10^{-3}$, $1 \times 10^{-1}$, $1 \times 10^{-4}$. The learning rate for the position of the Gaussians is exponentially interpolated between $3.2 \times 10^{-4}$ and $6.4 \times 10^{-6}$ based on the training steps, decreasing as the optimization progresses. In the supplementary material, a quantitative analysis of the impact of different learning rates demonstrate that our algorithm exhibits stability with respect to different choices of learning rates and is not sensitive to their variations within a reasonable range.

To accelerate the computation of the features, we truncate the opacity of the Gaussians when it is below a threshold $O_{threshold}$. For 3DGS objects, we set $O_{threshold}$ to $1/255$. For the ropes of the ship and flat Gaussian objects in GaMeS, to better represent the radiance field with spatial high-frequency information through the sampling points, we set $O_{threshold}$ to $1 \times 10^{-3}$. To accelerate the optimization, we update the coverage relationship between Gaussians and sample points every 10 steps. In practice, this approach yields results nearly identical with updating at every step.


Due to the presence of Bézier curves in the objects of the NeRF Synthetic Dataset, deforming them to create ground truth for comparison presents challenges. Therefore, we converted these objects to a mesh format and re-rendered the training and testing sets using Blender for our experiments. 

\begin{figure}[h]
    \centering
    \begin{subfigure}[b]{0.14\textwidth}
        \centering
        Input \& Deform
    \end{subfigure}
    \hfill
    \begin{subfigure}[b]{0.08\textwidth}
        \centering
        GT 
    \end{subfigure}
    \hfill
    \begin{subfigure}[b]{0.11\textwidth}
        \centering
        Baseline 
    \end{subfigure}
    \hfill
    \begin{subfigure}[b]{0.12\textwidth}
        \centering
        Ours 
    \end{subfigure}

    \begin{subfigure}[b]{0.48\textwidth}
        \centering
        \includegraphics[width=\textwidth]{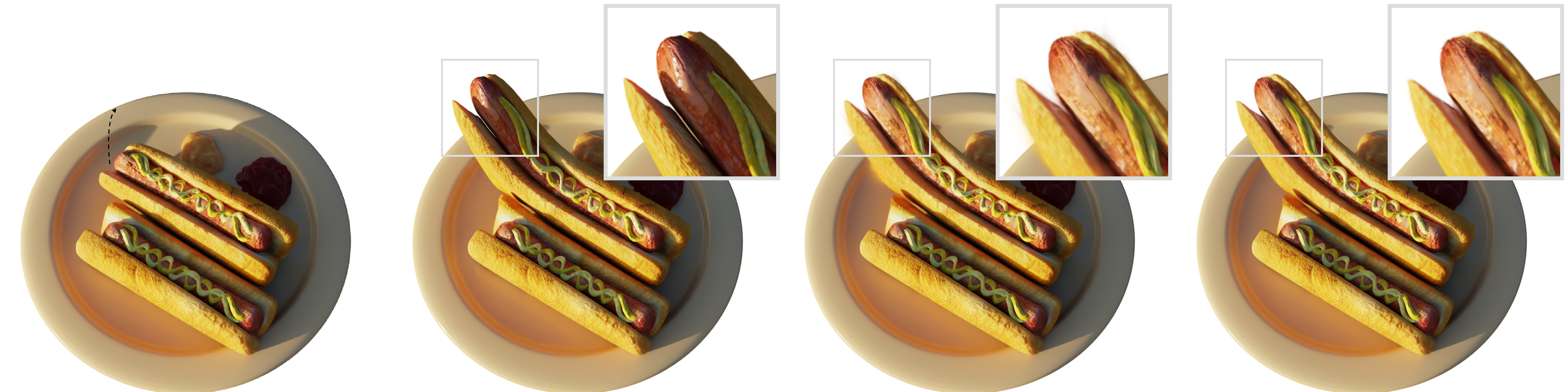}  
    \end{subfigure}

    \begin{subfigure}[b]{0.48\textwidth}
        \centering
        \includegraphics[width=\textwidth]{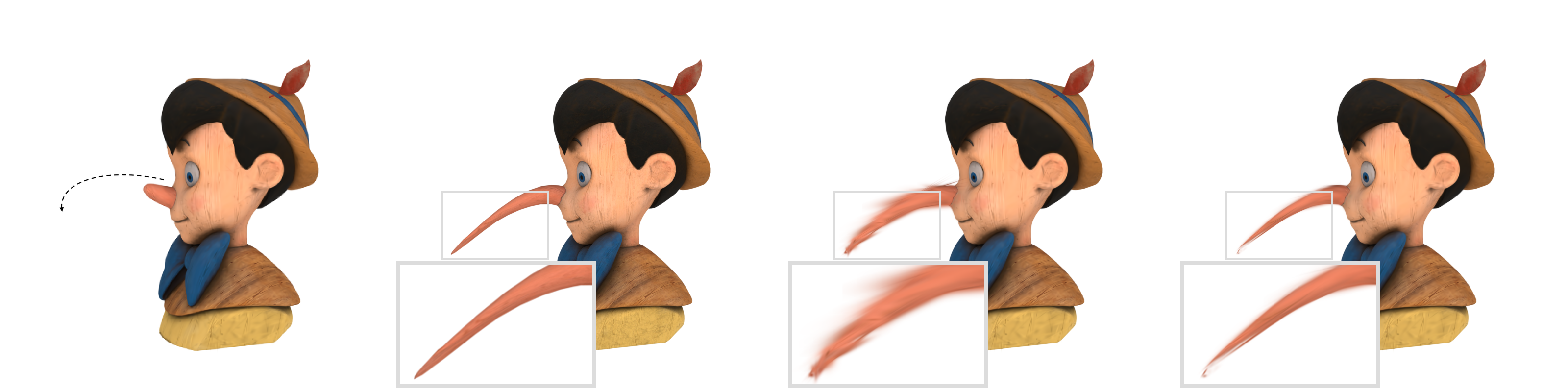}  
    \end{subfigure}

    \begin{subfigure}[b]{0.48\textwidth}
        \centering
        \includegraphics[width=\textwidth]{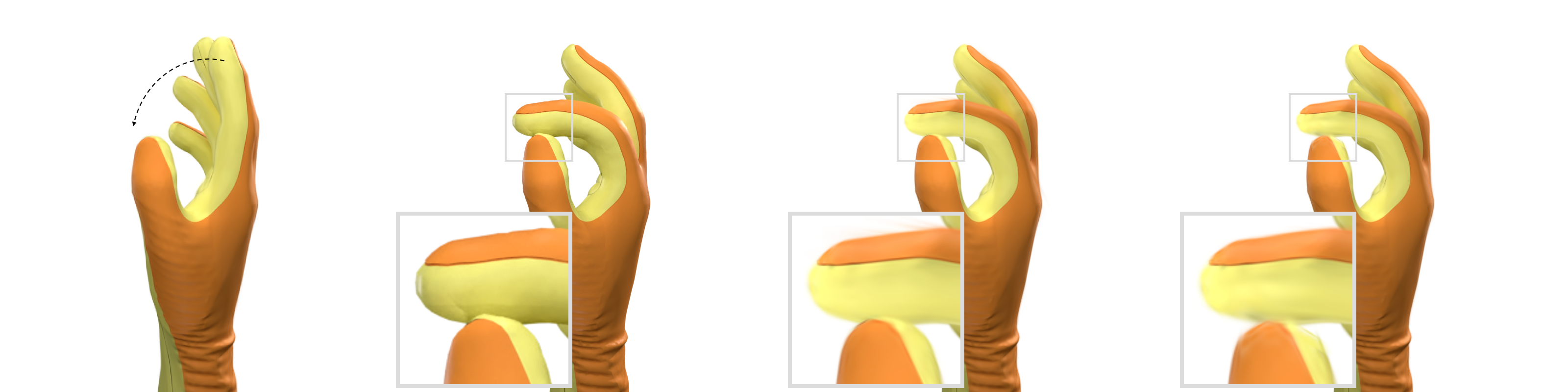}  
    \end{subfigure}

    \caption{\textbf{Failure cases.} The three rows respectively demonstrate that our algorithm fails to handle shadows and specular highlights, extreme deformations, and overlapping sampling regions within the deformed radiance field. }
    \label{fig:failed_cases}
\end{figure}

\subsection{More Deformation Results}
\label{app:more_results}
In Fig. \ref{fig:moreresults}, we present additional results of novel view synthesis under various types of deformations, including bending, twisting and scaling.

\section{Conclusion \& Future Work}
In this paper, we propose a novel method for performing ARAP deformation of Gaussian radiance fields. We are the first to introduce spatial features to describe the Gaussian radiance field for deformation. We also propose an approach that preserves the energy of the radiance field and optimizes Gaussians through this energy. We explore how spatial sampling during deformation can better maintain the radiance field and propose an adaptive anisotropic spatial low-pass filter. Our deformation algorithm does not rely on extracting geometric information from the Gaussian object and outperforms existing Gaussian editing techniques. We will open-source an interactive system that allows users to deform Gaussian radiance fields interactively.

In this work, we encapsulated all information, including the object's intrinsic color, material, environmental lighting, and occlusion, within a single radiance field. However, we did not account for color variations caused by occlusions, anisotropic materials, and lighting changes during deformation; therefore, we cannot properly handle shadows and the specular reflection component of the object during deformation. Since the radiance field is expressed through sampling, the number of sample points is constrained by memory capacity and processing speed, imposing an upper limit on the frequency of details that can be represented. Additionally, our framework may fail to handle extreme deformations or overlapping sampling regions in the deformed Gaussian radiance field. The failure cases are shown in Fig. \ref{fig:failed_cases}. In future work, we aim to explore methods to decouple lighting information within the Gaussian-based radiance field, seek continuous energy terms for radiance field representation, and integrate our framework with the 2DGS~\cite{HuangYC0G24} to obtain surface normals, which can be used to calculate the normal consistency with the groundtruth to validate the results. To facilitate further research and reproducibility, we provide our implementation at https://github.com/XinhaoT/ARAP-Deformation-of-Gaussian-Radiance-Fields.git.

\ifCLASSOPTIONcaptionsoff
  \newpage
\fi

\bibliographystyle{IEEEtran}
\bibliography{arap}

\appendices



\begin{IEEEbiography}[{\includegraphics[width=1in,height=1.25in,clip,keepaspectratio]{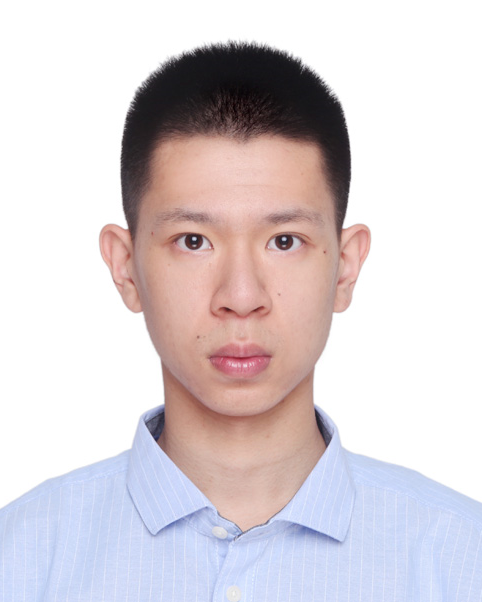}}]{Xinhao Tong} received his dual Bachelor's degree in Computer Engineering from Zhejiang University and the University of Illinois at Urbana-Champaign. He is currently a PhD student at Zhejiang University's State Key Lab of CAD\&CG, with research interests in 3D object reconstruction and editing.
\end{IEEEbiography}

\begin{IEEEbiography}[{\includegraphics[width=1in,height=1.25in,clip,keepaspectratio]{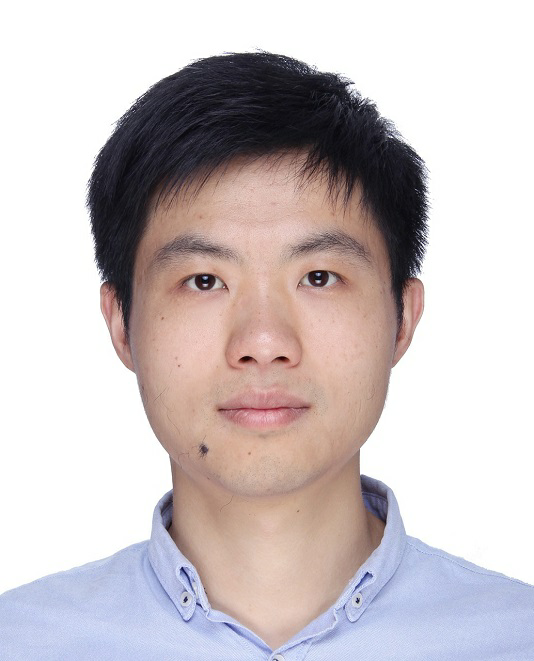}}]{Tianjia Shao} received his BS from the Department of Automation, and his PhD in computer science from Institute for Advanced Study, both in Tsinghua University. He is currently a ZJU100 Young Professor in the State Key Laboratory of CAD\&CG, Zhejiang University. Previously he was an Assistant Professor (Lecturer in UK) in the School of Computing, University of Leeds, UK. His current research focuses on 3D modeling from consumer hardware, and structure/function aware geometry processing.
\end{IEEEbiography}

\begin{IEEEbiography}[{\includegraphics[width=1in,height=1.25in,clip,keepaspectratio]{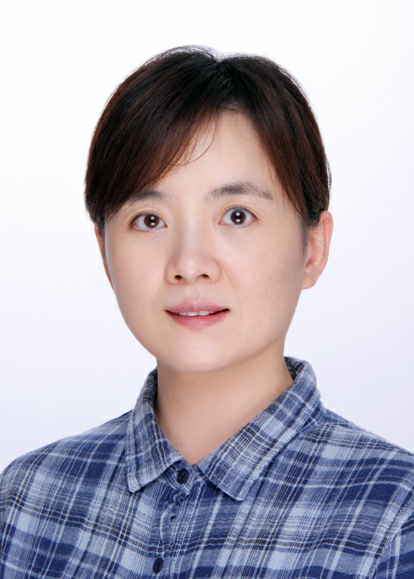}}]{Yanlin Weng} received the bachelor’s and master’s degrees in control science and engineering from Zhejiang University, and the PhD degree in computer science from the University of Wisconsin - Milwaukee. She is currently an associate professor with the School of Computer Science and Technology, Zhejiang University. Her research interests include computer graphics and multimedia.
\end{IEEEbiography}

\begin{IEEEbiography}[{\includegraphics[width=1in,height=1.25in,clip,keepaspectratio]{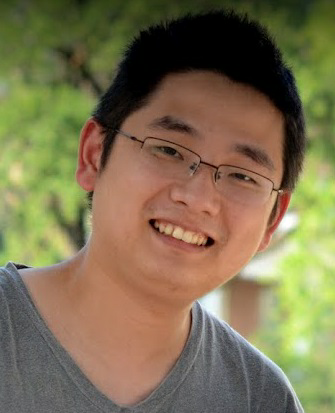}}]{Yin Yang}
received the PhD degree in computer science from the University of Texas at Dallas, in 2013. He is an associate professor with Kahlert School of Computing, University of Utah. He co-direct Utah Graphics Lab with his colleague Prof. Cem Yuksel. He is also affiliated with Utah Robotic Center. Before that, He was a faculty member at University of New Mexico and Clemson University. His research aims to develop efficient and customized computing methods for challenging problems in Graphics, Simulation, Deep Learning, Vision, Robotics, and many other applied areas.
\end{IEEEbiography}

\begin{IEEEbiography}[{\includegraphics[width=1in,height=1.25in,clip,keepaspectratio]{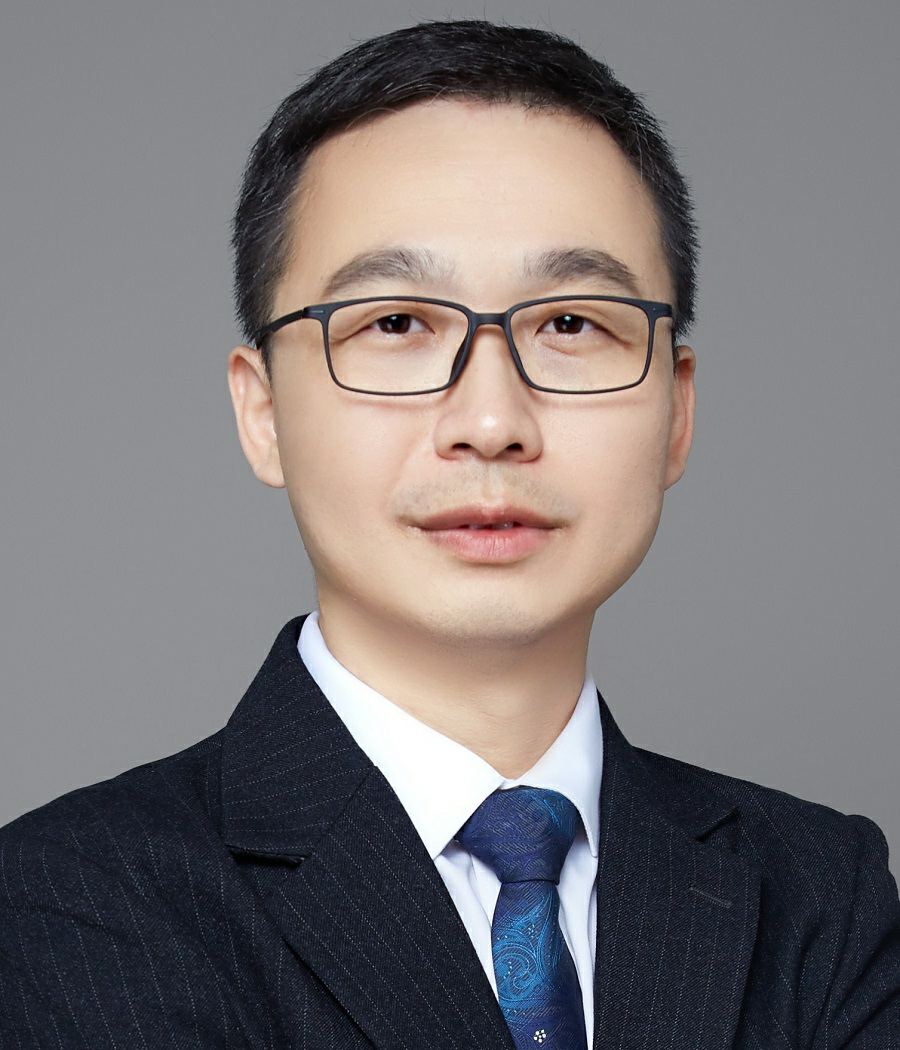}}]{Kun Zhou}
is a Cheung Kong Professor of Computer Science at Zhejiang University and the Director of the State Key Lab of CAD\&CG. He received his PhD degree from Zhejiang University. After graduation, he spent six years with Microsoft Research Asia, serving as a lead researcher of the graphics group before returning to Zhejiang University. He was named one of the world's top 35 young innovators by MIT Technology Review (2011), and received the Asiagraphics Outstanding Technical Contributions Award (2022) and the ACM SIGGRAPH Test-of-Time Award (2024). He is a Fellow of IEEE and ACM.
\end{IEEEbiography}





\end{document}